\documentclass[%
 amsmath,amssymb,
 aps,
 prb,
 twocolumn
]{revtex4-2}

\usepackage{graphicx}
\usepackage{dcolumn}
\usepackage{bm}
\usepackage{xcolor}
\usepackage[caption=false]{subfig}
\usepackage{tikz}
\usepackage{xfrac}
\usepackage{blindtext}
\usepackage{times}

\renewcommand{\vec}[1]{{\boldsymbol #1}}

\definecolor{darkblue}{HTML}{004D6B}
\definecolor{darkred}{HTML}{8c1515}

\usepackage{hyperref}
\hypersetup{
	colorlinks=true,
	urlcolor=darkred,
	citecolor=darkblue,
	linkcolor=darkred,
	breaklinks
}

 \newcommand{\moire}{moir\'e\ }

\begin{document}

\title{Universal principles of moir\'e band structures}

\author{Jan Attig}
\thanks{These two authors contributed equally to this work.}
\author{Jinhong Park}
\thanks{These two authors contributed equally to this work.}
\author{Michael M. Scherer}
\author{Simon Trebst}
\author{Alexander Altland}
\author{Achim Rosch}
\affiliation{Institute for Theoretical Physics, University of Cologne, 50937 Cologne, Germany}

\date{\today}

\begin{abstract}
	Moir\'e materials provide a highly tunable environment for the realization of  band structures with engineered physical properties. Specifically,  \moire structures with Fermi surface flat bands --- a synthetic environment for the realization of correlated phases --- have \moire unit cells containing thousands of atoms and tantalizingly complex bands structures. In this paper we show that \emph{statistical principles} go a long way in explaining universal physical properties of these systems. Our approach builds on three conceptual elements: the presence of quantum chaos caused by the effective irregularity of the atomic configurations on short length scales, Anderson localization in momentum space, and the presence of approximate crystalline symmetries. Which of these principles dominates depends on material parameters such as the extension of the Fermi surface or the strength of the \moire lattice potential. The phenomenological consequences of this competition are predictions for the characteristic group velocity of \moire bands, a primary indicator for their average flatness. In addition to these generic features, we identify structures outside the statistical context, notably almost flat bands close to the extrema of the unperturbed spectra, and  the celebrated zero energy `magic angle' flat bands, where the latter require exceptionally fine tuned material parameters.
\end{abstract}

\maketitle


\section{Introduction}
Sheets of two-dimensional materials stacked at relative twist angles or with a
mismatch in lattice constant define a class of quantum matter known as \moire
materials. At low twist angles or small lattice constant mismatch, \moire materials
can have tens of thousands of atoms in their effective unit cells, and as many energy
bands in their Brillouin zones. Controlled variations of twist angles, and/or the
(corrugated) van der Waals coupling between layers affords the unique opportunity of
band structure
engineering~\cite{hunt2013massive,macdonald2011graphene,bistritzer2011moire,yankowitz2018dynamic,rode2017twisted,yankowitz2012emergence,balents2020superconductivity,can2021high}.
The recent realization of almost non-dispersive bands in twisted bilayer graphene (TBG)
(and the observation of a wealth of strong correlation
effects~\cite{cao2018correlated,cao2018unconventional,yankowitz2019tuning,choi2019electronic,lu2019superconductors,kerelsky2019maximized,jiang2019charge}
symptomatic for flat band materials)  demonstrate  the opportunities provided by this
type of quantum matter, which besides
graphene~\cite{chen2019signatures,chen2019evidence,liu2020tunable,shen2019observation,burg2019correlated,he2020symmetry}
contains hexagonal boron nitride~\cite{xian2019multiflat,ni2019soliton}, transition
metal dichalcogenides~\cite{regan2020mott,wang2020correlated,tang2020simulation}, and
others~\cite{xian2020realization,kennes2020moir} as material platforms.

\begin{figure}[b]
\centering
\includegraphics[width=\columnwidth]{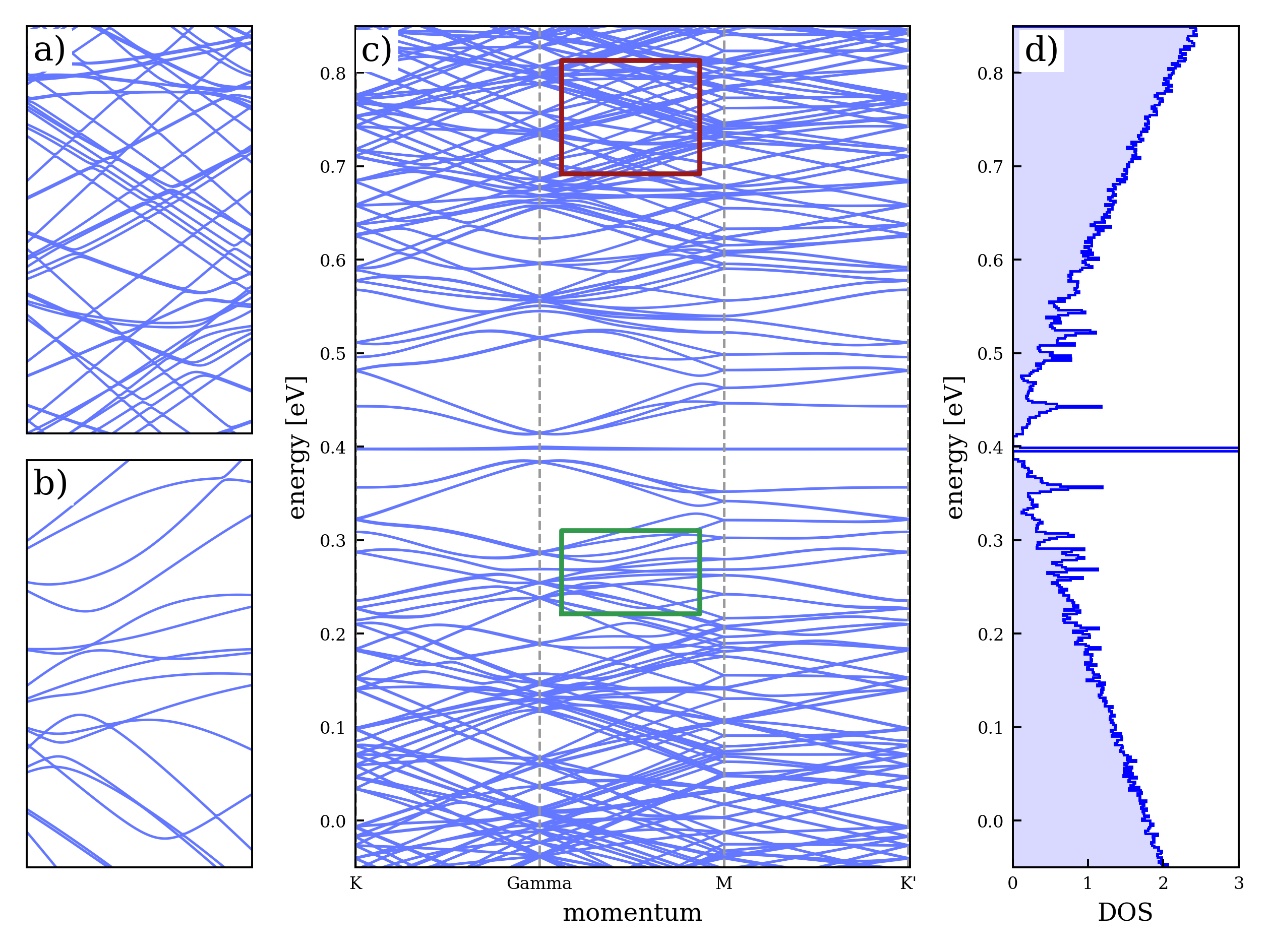}
\caption{{\bf Features in the band structure of moir\'e materials}.
		Shown is  a typical band structure of twisted bilayer graphene, with zooms
		into the band structure (c) and two cutouts (a-b) as well as density of states
		(d) for two layers of graphene twisted by a commensurate angle of $\theta
		\approx 1.1^\circ$.		
		}
\label{Fig:TBG}
\end{figure}

The tantalizingly complex band structure of \moire materials raises the question for
underlying universal principles. For instance, a naked eye inspection of the blow-ups
in Fig.~\ref{Fig:TBG}  reveals recurrent patterns in the `spaghetti' of individual
energy bands. Most apparent among these are regions with almost linearly dispersive
bands (`uncooked spaghetti' in Fig.~\ref{Fig:TBG}(a)) interspersed by narrowly
avoided crossings, and regions with slack energy bands (`cooked spaghetti' in
Fig.~\ref{Fig:TBG}(b)) meandering up and down subject to strong band repulsion. The
average uniformity of these patterns over wide ranges of momenta and energies
suggests that statistical principles are at work. Embedded in these structures we
observe anomalous features which clearly are not of statistical nature. Most
prominent among these are the celebrated flat bands forming upon fine tuning of twist
angles and/or corrugation parameters, see Fig.~\ref{Fig:TBG}(c) near the Fermi level
of $\sim$\,0.4\,eV. In addition to these exceptional flat bands, there are more
robust `super-flat' bands forming next to the band minima of the uncoupled layers, as
we will show below.

In this paper, we present a simple semi-phenomenological theory which turns the complexity implied by the large number of bands into an advantage and uses it as the basis for a \emph{statistical approach}~\cite{10.2307/1970079,guhr1998random,mehta2004random}. Our theory builds on  three basic principles:
\begin{itemize}
	\item \emph{Lattice periodicity}. The presence of an effective \moire potential repeating itself over large distance scales defines a lattice structure of periodicity $L\gg 1$ atomic lattice spacings, $a$. The latter may be looked at in real space or, preferably for our purposes, in momentum space. In that representation, the system is described by a finite lattice of spacing $G_m\sim 1/(La)$ with $\mathcal{O}(L^2)$ sites corresponding to the number of atoms in the \moire unit cell. The \moire potential defines an effective hopping Hamiltonian in this lattice, and the dispersion relation of the unperturbed layers that of an effective on-site potential.
	\item \emph{Anderson Localization}. The aperiodic site-to-site variations of that potential define a source of effective irregularity or quantum \emph{disorder}. For weak \moire potentials, hopping along the effectively one-dimensional equi-potential Fermi surfaces is impeded by the mechanism of Anderson localization (in momentum space). We will see that this manifestation of quantum localization is an efficient promoter of strong energy band dispersion. 
	\item \emph{Quantum chaos}. With increasing \moire hopping the Fermi surfaces broaden and eventually turn into quasi two-dimensional structures. The increasing hopping strength driving this development delocalizes  wave functions, up to a point that they cover a two-dimensional subset of the lattice almost ergodically. In such regimes, the combination of residual quantum disorder and discrete symmetries characterizing the \moire lattice defines weakly dispersive band structures containing accumulations of almost, but not fully flat bands. 

\end{itemize}
As we will see, the combination of these three elements goes a long way in quantitatively describing the universal features of the \moire band structure. However, it also explains  various non-universal features, among them the formation of different types of flat bands or van Hove structures.

\section{Setup and general considerations}
\label{sec:setup}

\begin{figure}
	\centering
	\includegraphics[width=\columnwidth]{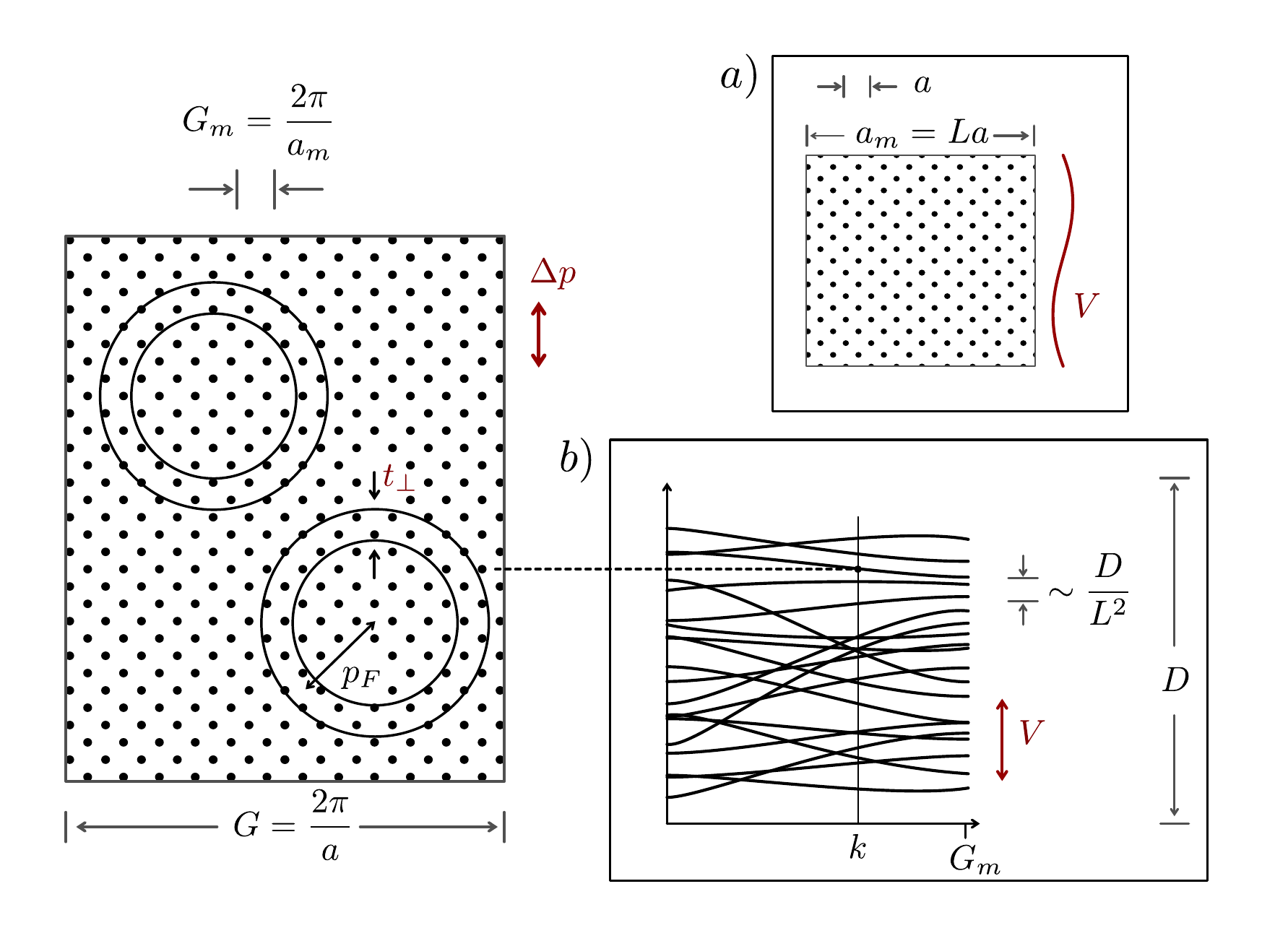}
	\caption{{\bf Extended Brillouin zone of a \moire lattice.} A \moire potential $V$ periodic in real space over scales $La\equiv a_m\gg a$, cf. inset a),  has two interrelated effects: it defines a \moire Brillouin zone with reciprocal lattice vector magnitude $G_m=2\pi/a_m$ and $\mathcal{O}(L^2)$ bands of characteristic spacing $\sim D/L^2$, where $D$ is the total bandwidth, cf. inset b). Second, it causes hopping with characteristic strength $t_\perp 
	\sim V$ between momentum states differing by multiples of the \moire reciprocal lattice vectors in the extended Brillouin zone (left). In this way we obtain, for each fixed instance $k$ of the conserved \moire cell momentum, a lattice of momentum states subject to short range hopping with characteristic momentum transfer $p \sim G_m\ll G$ and hopping rates $\sim V\sim t_\perp$,  typically satisfying the inequalities $D\gg V\gg D/L^2$. For a given Fermi energy, $\epsilon_F$, wave functions subject to the competition of potential and hopping are generically confined to the broadened Fermi surfaces discussed in the text.}
	\label{Fig:MoireBandStructure}
\end{figure}

We consider two-dimensional crystalline systems with lattice constant $a$ -- more
generally $(a_1,a_2)$ -- subject to a perturbation $V$ periodic over distances
$a_m=La$, $L\gg 1$ (inset a) of Fig.~\ref{Fig:MoireBandStructure}). This perturbation may  be the substrate  potential induced by the proximity  of a second layer of
different chemical composition, or generated by the coupling to a second layer as in twisted bilayer graphene. In either case, $V$ defines periodic `hopping potential' in momentum space, and we discuss them in parallel. More specifically,  in the `extended' Brillouin
zone defined by the reciprocal lattice vector of magnitude $G=2\pi/a$, the potential
$V$ defines transition matrix elements between states differing in multiples of the \moire reciprocal
lattice vector of magnitude $G_m\equiv 2\pi/a_m=G/L$, and thus defines a momentum space hopping Hamiltonian. Momenta $\vec{k} \,
\mathrm{mod}\, \vec{G}_m$ are conserved, and on this basis, the spectrum of
the system gets organized into $\sim L^2$ bands indexed by $\vec{k}$. We assume the
potential to be strong enough to  couple these bands, $V\gtrsim D/L^2$,
where $D$ is the total bandwidth and $D/L^2$ defines the characteristic band spacing. This assumption rests on the efficient coupling of `sites' in the
momentum space \moire lattice over scales $\Delta p\gtrsim G_m$. 
(In Appendix
\ref{App:1D-Moire} we discuss in which way  this assumption relies on the  commensurability properties of the substrate potential and is not entirely innocent.)

In the rest of the paper, we will build on a momentum space lattice picture to obtain
information on  universal features of the \moire band structure. In particular, we
will emphasize connections between the present problem and that of Anderson
localization in quasi one-dimensional disordered media. To understand this link,
notice that for each realization of $\vec k$ the dispersion of the native
two-dimensional material $\epsilon_{\vec{Q}_n}\equiv \epsilon(\vec k + \vec{Q}_n) $ is a
function on the sites $\vec{Q}_n \equiv n_1 \vec{G}_{m,1}+ n_2
\vec{G}_{m,2}$ of the \moire lattice with effectively random site--to--site
variations and continuous modulation in $\vec{k}$.

 In momentum space, $\epsilon_{\vec{Q}_n}$ acts as
an \emph{effective potential}. In combination with weak (translationally invariant) hopping $t_\perp$ induced by $V$, this potential has two principal effects. First, it defines broadened \emph{quasi one-dimensional} Fermi surfaces, centered around the Fermi lines $\epsilon_{{\vec{Q}_n}}=\epsilon_F$ of the unperturbed system. More precisely, typical  wave functions will  probe a region of width $\sim t_\perp / (v_F G_m)$ lattice spacings around these contours, see Fig.~\ref{Fig:MoireBandStructure}. Second, as the reciprocal lattice is discrete, the site-to-site variations of  $\epsilon_{\vec{Q}_n}$, define a source of effective disorder.  

Such \emph{quasi} one-dimensional disordered systems are subject to Anderson localization weaker than in strictly one-dimensional materials (potential roadblocks can be efficiently sidestepped) but stronger than in two-dimensions (there is limited phase space for transverse diffusion).
We aim to explore how the ensuing physics of wave function confinement due to effective disorder manifests itself in the universal band structure of \moire materials and what exceptions to such a universal framework exists.

To sharpen the question, let us for a moment ignore the localization principle and
speculate on  the ramifications of the coupling $V$ in the band structure. Under the
above assumptions, $V\gtrsim D/L^2$ is a strong and effectively random perturbation
parametrically dependent on $\vec k$. On this basis, we should expect efficient level
repulsion, i.e. variations of bands over scales $\Delta \epsilon \sim D/L^2$ under
parametric variations of extension $\Delta k \sim G_m$  --- the `cooked spaghetti'
scenario. Characteristic level `velocities' in this case would be of order $\Delta
\epsilon /\Delta k\sim v_F (a/L)^\alpha$, where $v_F \sim D/a$ is a typical group velocity of the underlying two-dimensional material and the exponent $1/2 \le \alpha \le 1$ will depend on how strongly neighboring `cooked spaghetti' will wiggle relatively to each other, see below. Regions of such small dispersion are observed for sufficiently large $t_\perp$, but they
are not generic. Far more frequently do we see `uncooked spaghetti', of steeper
velocity $\mathcal{O}(L^0)$. This is proof--by--contradiction that `quantum disorder' or `quantum chaos' reasoning by itself does not explain the generic band structure. We aim to demonstrate how the localization principle is the missing element in the story.

While much of our phenomenological reasoning does not rely on model specific assumptions, the concrete calculations below are performed for the case of twisted bilayer structures, more specifically, bilayers of honeycomb lattices as relevant to the case of magic-angle graphene. The details of this model setup are summarized in Appendix~\ref{sec:TwistedBilayerGraphene}.

In this paper, we will look at this system through the lenses of two complementary numerical models. The first is a real space tight-binding model of twisted honeycomb layers with Slater-Koster parameters~\cite{PhysRev.94.1498,de2009localization,PhysRevB.87.205404}, cf.~App.~\ref{app:realspace}. This model describes the system from first principles and is suited to explore wide portions of its spectrum, including those far detached from its Dirac points. However, this freedom comes at the expense of relatively high computational demands. We use this approach to explore large scale statistical features of the spectrum. The second is a  continuum model developed by MacDonald and
Bistritzer for bilayer graphene \cite{bistritzer2011moire} (see
App.~\ref{app:momentum} for a review). This model assumes linear dispersion of the uncoupled layers, and hence is limited to the vicinity of the Dirac points. However, due to its computational efficiency it gives us highly resolved insight into the spectra and wave functions near these points.

The rest of the paper is organized as follows: In section~\ref{sec:Waterfalls} we
discuss  key  observations on the band structures of \moire
materials obtained within the framework of the real space approach. 
In section~\ref{sec:MomentumSpaceLocalization} we introduce
the real and momentum space description of these structures as a basis for our
subsequent discussion of the statistical approach. 
In the central section~\ref{sec:three_different_regimes} we discuss how the band velocity distributions relates to the concept of Anderson localization in momentum space. This is followed by the detailed numerical study of the continuum model in section~\ref{sec:ContinuumModel} where we consider spectral and wave function statistics as indicators of chaos and localization, and relate them to the observed characteristics of the velocity statistics.  We conclude in section~\ref{sec:discussion}, technical details of our analysis are relegated to several appendices.


\section{Statistics of velocity distribution}
\label{sec:Waterfalls}

\begin{figure}[t]
	\centering
	\includegraphics[width=\columnwidth]{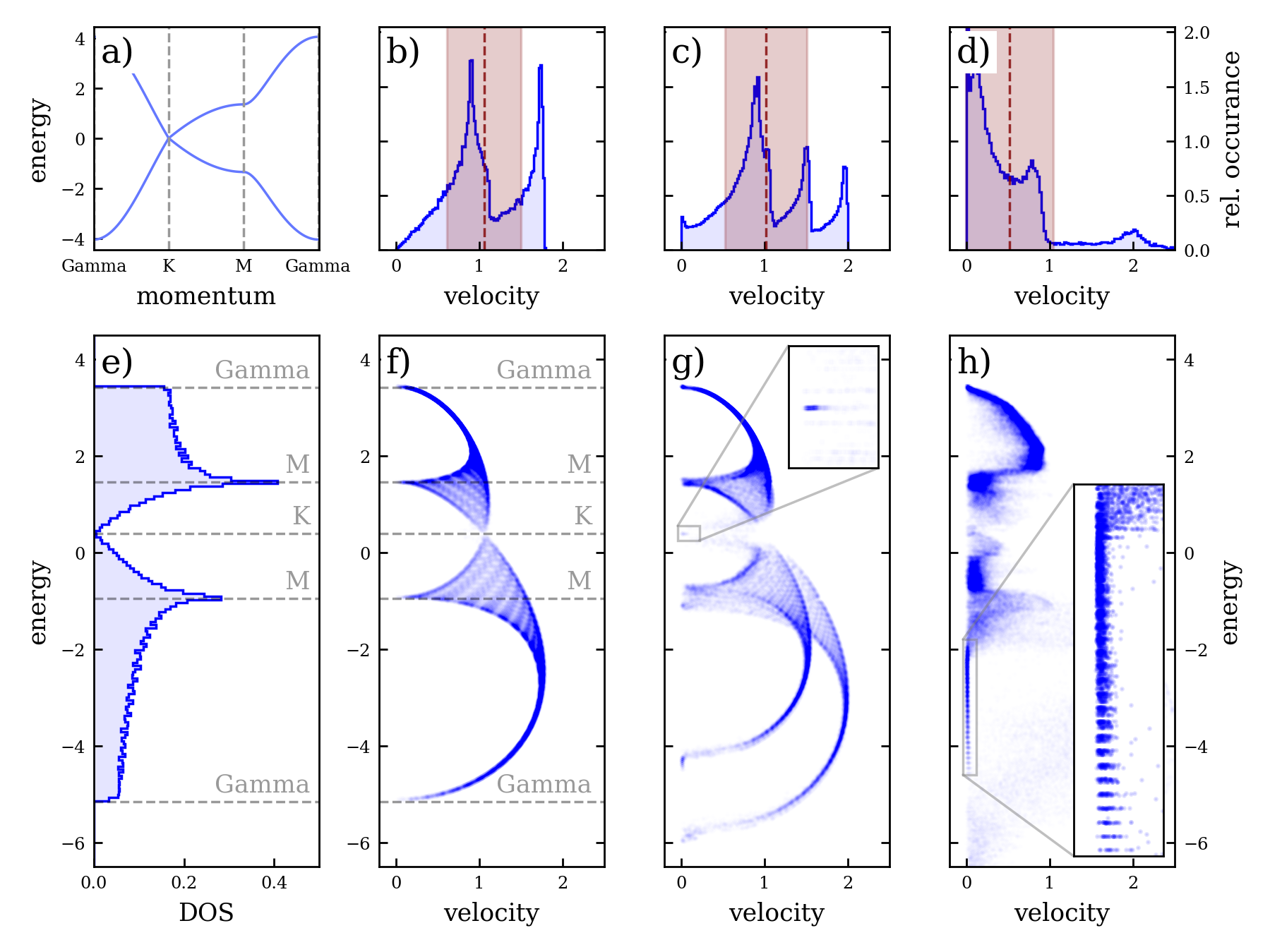}
	\caption{{\bf Velocity distribution in TBG model.}
	Variations in velocity distributions (b-d) and velocity-energy correlations (f-h) when increasing interlayer coupling for two graphene layers at commensurate angle of $\sim 1.1^\circ$, (corresponding to $m=27, n=26$, cf. App.~\ref{app:realspace} for a definition). 
	Panels b), e), and f) show data for decoupled and uncorrugated layers of graphene, i.e. $V = 0$, $C=0$.
	Panels c) and g) show data that is compatible with experimental studies, i.e. $V=1$,$C=1$.
	The emerging flat band can be seen as a bright spot in the velocity-energy correlations and is highlighted in panel g).
	Panels d) and h) show data for an extremely corrugated and strong layer-coupled case with parameters $V=2$, $C=5$.
	Harmonic oscillator ladder states emerge for low energies, highlighted in the inset of panel h).
	For comparison, panel a) shows the dispersion of nearest-neighbor coupled single-layer graphene.}
	\label{Fig:Waterfalls}
\end{figure}

To set the stage for our analysis, we first discuss the band structure of the twisted honeycomb bilayer \moire system,
with an emphasis on the distribution of \emph{band velocities} over wide ranges of the spectrum.
Our method of choice in this endeavor is the real space model~\cite{PhysRev.94.1498,de2009localization,PhysRevB.87.205404} mentioned above. 
We diagonalize the model to extract the dispersion of the energy
bands~$\epsilon_n(\vec{k})$ with band index $n$ and wave vectors $\vec{k}$ within the
first moir\'e Brillouin zone. The  electron group velocity with wave vector $\vec{k}$
in band $n$ is then given as
$\vec{v}_n(\vec{k})=\nabla_{\vec{k}}\epsilon_n(\vec{k})$. To obtain the velocity
distribution, we collect the absolute values $|\vec{v}_n(\vec{k})|$ in all bands for
a large set of randomly selected wave vectors. We show their relative occurrence in
the right three panels of the upper row of Fig.~\ref{Fig:Waterfalls} for three
different choices of the interlayer coupling and corrugation, ranging from decoupled
flat layers (c.f.~panels b+f) over the experimental system (c.f.~panels c+g) to
strongly coupled and corrugated layers (c.f.~panels d+h). The corresponding panels in
the lower row show which velocities appear at a given energy.

We first observe that typical electron velocities are of order $v_{\mathrm{F}}$ (cf.~the vertical dashed lines in panels~b) and c) in Fig.~\ref{Fig:Waterfalls} for average values), and hence do not suffer the suppression down to scales $\sim 1/L$ which would be expected on the basis of the naive `band repulsion' picture formulated above.
On the other hand, the distributions tend to exhibit enhanced probabilities for
smaller velocities upon increasing corrugation and interlayer coupling, cf.~panel~d)
in Fig.~\ref{Fig:Waterfalls} where the interlayer tunneling is twice as large as
experimentally reported and the corrugation is increased by a factor of five. In this
extreme parameter regime, the main peak of the velocity distribution is shifted
towards small velocities right above zero. The corresponding panel~h) indicates that
the small velocities come from different energy regions at and near the van-Hove
filling. 

In addition to the generic features of the
energy--velocity distribution, we observe a number of anomalies in the form of almost
perfectly non-dispersive (zero velocity) bands. Among these, the most prominent is
the celebrated flat band of twisted bilayer graphene, visible as a bright spot in the
inset of panel g). The interpretation of this anomaly in the mindset of the present approach is discussed in section~\ref{sec:discussion}.
At strong corrugation we observe a different type of flat bands, distinguished by their uniform spacing in energy (inset of panel h). This ladder structure affords a natural interpretation as a tunneling phenomenon in the \moire momentum space lattice structure, as we discuss in Appendix \ref{app:ladders}. Here, we will focus on the analysis of the `generic' regions of the velocity distribution and their explanation in terms of momentum space localization.

\section{Momentum space localization}
\label{sec:MomentumSpaceLocalization}

With an eye on the generic regions of \moire band structures, we now proceed to develop the momentum space picture in detail 
and how it can be used to characterize the universal features of  \moire band structures.
For the sake of definiteness, we again consider the case of twisted bilayer graphene.
However, most of the discussion applies to different types of \moire materials with little or no alteration.

We consider a lattice in its momentum space representation, as
schematically depicted in the inset of Fig.~\ref{Fig:MomentumSpaceLocalization}(a).
General wave functions $|\psi_{n} (\vec{k}) \rangle$ with momentum $\vec{k}$ in the
first \moire Brillouin zone and band index $n$ are given by
\begin{align}
|\psi_{n} (\vec{k}) \rangle = \sum_{\vec{Q}} \sum_{\alpha = U, L} A_n^{\alpha} (\vec{k} - \vec{Q}) | \psi_\alpha (\vec{k} -  \vec{Q} - \vec{K}^{\alpha})  \rangle \,.
\end{align}
Here $\vec{Q}$ are the \moire reciprocal lattice vectors, $\alpha$ is the layer index ($\alpha = U, L$), $\vec{K}^{\alpha}$ is the $K$ point of layer $\alpha$ in the first \moire Brillouin zone, cf. Fig.~\ref{Fig:MoireBandStructure}(b), $A_n^{\alpha} (\vec{k} - \vec{Q})$ are expansion coefficients, and $|\psi_{\alpha} (\vec{k} +  \vec{Q}  - \vec{K}^{\alpha}) \rangle$
are two-component vectors whose components
describe the wave function amplitudes on the $A$ and $B$ sites of the bipartite honeycomb lattice, respectively.
Using $|\psi_{n} (\vec{k}) \rangle$ as basis vectors, one can evaluate the tunneling matrix elements between layers and derive an effective Hamiltonian, see Appendix \ref{app:momentum}.

Hereafter, we concentrate on physics near the $K$ points for simplicity.
By folding out of the first \moire Brillouin zone to the extended zone, a lattice  spanned by the reciprocal \moire lattice vectors  $\vec{Q}$  (cf. gray dots in Fig.~\ref{Fig:MomentumSpaceLocalization}) emerges. 

For the following discussion it is useful to simplify our setup still further and temporarily ignore the layer and the $K$-point index. The essential physics is then described by the Hamiltonian
\begin{align}
	\label{eq:LatticeHamiltonian}
	\hat H_{\vec k} = -t_\perp\! \sum_{\langle \vec Q,\vec Q'\rangle} (c^\dagger_{\vec{Q}}c^{\vphantom{\dagger}}_{\vec{Q}'}+\text{h. c.})+\sum_{\vec{Q}}\epsilon(\vec k +\vec{Q})c^\dagger_{\vec{Q}}c^{\vphantom{\dagger}}_{\vec{Q}} \,.
\end{align}
where $c^\dagger_{\vec Q}$ creates a particle at momentum $\vec k+\vec Q$ and we made the parametric dependence on the conserved momentum $\vec k$ explicit. 

As discussed above, the Hamiltonian
Eq.~\eqref{eq:LatticeHamiltonian} describes the spreading of wave functions along
contours of constant energy $\epsilon_{\vec{Q}}\simeq \epsilon=\,\text{const}.\,$.

A wave function hybridizes over two nearest neighbors of the momentum lattice on these shells provided the energy difference  is of order $\Delta \epsilon \sim v_F G_m \lesssim t_\perp$.
This hybridization criterion gives the shells a width of $\sim t_\perp/v_F G_m$ in \emph{transverse} direction. In the \emph{longitudinal}  direction, the effectively random site energy variations make quasi one-dimensional localization an inevitable consequence.

However, the question
remains under what conditions that localization length $\xi$ is smaller or larger
than the circumference of the momentum space Fermi ring. Unfortunately, finding
parametric estimates for the dependence
$\xi(t_\perp,\epsilon,\{\epsilon_{\vec{Q}}\})$ of the localization length on the
relevant system parameters is not easy under the present circumstances. The reason is
that for most systems of interest, $t_\perp$ is of the same order 
as the characteristic energy differences $\delta \epsilon\sim
|\epsilon_{\vec{Q}}-\epsilon_{\vec{Q'}}|$ between nearest neighbors, i.e. $t_\perp
\simeq \delta \epsilon$. We are thus sitting between the two chairs of localization
in  strongly and weakly disordered media, respectively. On top of that, the lattice
contains stretches of sites approximately aligned with the Fermi surface (see
Appendix~\ref{app:incomm} for further discussion and illustration). Along these,
site-to-site energy differences are atypically small, defining local corridors of
near ballistic wave function propagation.

\section{Three localization regimes}  
\label{sec:three_different_regimes}

The momentum space setup introduced above sets the stage for the identification of three different regimes of qualitatively different phenomenology. 
The nature of these is best understood by considering what happens as the interlayer coupling, $t_\perp$, is gradually increased for a system at given Fermi energy~$\epsilon$:

\subsubsection*{I: Deep localization regime} 
\label{ssub:i_deep_localization_regime}

 For small site hopping, $t_\perp$, we are in a regime of strong Anderson localization. Individual eigenfunctions, $\psi_n$ are centered around specific momenta $\vec k_n =\langle \psi_n|\hat{\vec{k}}|\psi_n \rangle$ of magnitude $k_n\sim \epsilon/v_F$. 
The group velocity of the electron is computed from $\langle \psi_n|\frac{\partial H_{\vec k}}{\partial \vec k}|\psi_n \rangle$ and changes very little as long as the wavefunction is localized in close proximity to $\vec k_n$. At the same time, $|\partial_{\vec k}H_{\vec k}|\sim v_F$ is large, giving the  band dispersion the structure of steep almost linear functions of $\vec k$ --- the regime of uncooked spaghetti.   We finally notice that the complete momentum space localization implies the absence of correlations between distinct eigenfunctions and their eigenvalues, with the observable consequence of Poissonian spectral statistics. This regime is of direct relevance for bilayer graphene at large twist angle, see below.

\subsubsection*{II: Quasi one-dimensional (de-)localization} 
\label{ssub:ii_incomplete_localization}

 An increasing of $t_\perp$ causes a gradual compromising of the pristine momentum space localization. Wave functions begin to spread around the Fermi ring, and thus become correlated. While analytic computations are not straightforward, our numerical analysis below suggests an approximately linear dependence, $\xi\sim  (t_\perp/v_F)$ for the localization length in momentum space for small $t_\perp$. Assuming that in the same regime the radius of the Fermi surface $\sim \epsilon/v_F$ is linear in the Fermi energy, we obtain a crossover scale $t_\perp \sim \epsilon$ for the I/II regime boundary. 
Inside regime II, delocalization combined with the presence of effective randomness should lead to chaotic (Wigner-Dyson) correlations in the energy spectra. At the same time, individual states remain inhomogeneously distributed, with non-vanishing expectation values of magnitude $k_n \sim \epsilon/v_F$. On this basis, we expect bands with continued steep slope, but showing the `level repulsion' symptomatic for chaotic spectra. Metaphorically, this is a regime of semi-cooked spaghetti.

\subsubsection*{III: Strong coupling and dimensional crossover} 
\label{ssub:iii_strong_coupling_regime_}

 Naively, one would expect that  further increase of $t_\perp$ results in an
\emph{ergodic} phase characterized by uniform wave function distribution around the
Fermi surface and strong level correlations. However, the actually observed behavior
is more nuanced. In fact, the increase of the coupling, starts several developments,
the confluence of which determines the observable phenomenology: for coupling
strength approaching $t_\perp \sim \epsilon$, wave functions are no longer confined
to a ring, they flood the interior of the constant `potential' energy circle, and
significantly extend beyond it. We note that for models with Dirac dispersion this criterion for the II/III boundary
is parametrically of the same order than that for the I/II boundary, indicating that
the intermediate regime II may not have a parametrically wide support. On the other hand, the three regimes are defined by physically different principles and  our
analysis below demonstrates the prevalence of regime II over a numerically wide interval for
relevant model parameters. 

Second, the diminishing influence of the effectively random fluctuations in
$\epsilon_{\vec{Q}}$ implies a higher degree of wave function isotropy. For the same
reason,   discrete \emph{symmetries} begin to play a role. The \moire momentum space
lattice shows crystal discrete symmetries, namely $C_3$ rotation symmetry and the
mirror symmetry $M_y$ ($y \rightarrow -y$) combined with an operation that flips the
upper and lower layers. Except at few high symmetry points in the reduced
$\vec{k}$-Brillouin zone, these symmetries are broken by the parametric momentum
$\vec{k}$ in the function $\epsilon_{\vec{Q}}=\epsilon_{\vec{Q}}(\vec{k})$. However
for strong  $t_\perp$ this symmetry breaking gets relatively weaker, especially in
the neighborhood of high symmetry points. In principle, one may expect a situation
where the Hilbert space is reorganized into irreducible symmetry representation
spaces of these, with chaotic correlations inside each symmetry sector due to the
residual influence of $\epsilon_{\vec{Q}}$ and only weak correlations between
different sectors. This anticipation suggests spectral statistics intermediate
between Wigner-Dyson and Poissonian. At the same time, we expect band velocities
parametrically smaller than in regime II. To be a little more concrete, assuming that
a characteristic wave function spreads over $N$ lattice sites exploring both positive
and negative velocities.  In this case, the central limit theorem suggests a velocity
expectation value of order $v_F /\sqrt{N}$ with $N \sim L^2$ in the two-dimensional
regime while $N\sim L$ in the quasi one-dimensional regime. While this estimate may
be too crude, our analysis below  confirms hybrid Wigner-Dyson/Poisson hybrid
statistics and band velocities drastically reduced compared to those in regime II.

\section{Continuum Model}
\label{sec:ContinuumModel}

From our previous discussion it is evident that both the structure of single wave
functions and correlations in the energy spectrum play a crucial role in 
understanding the physics of the above regimes I-III. On this basis, we focus on two sets of statistical observables throughout: (i) wave-function statistics,
as characterized by  inverse participation ratios, and (ii) spectral statistics
described by the so-called Kullback-Leibler divergence. The latter is particularly suited to the quantification of spectral statistics in hybrid regimes where neither Poisson nor Wigner-Dyson statistics prevails in pure form. We will analyze these
quantities within the framework of a continuum model developed by MacDonald and
Bistritzer for bilayer graphene \cite{bistritzer2011moire} (see
App.~\ref{app:momentum} for a review). Focusing on the strong corrugation limit
\footnote{ We focus on a strong corrugation limit where the interlayer coupling in
the AA region is zero since in the weak corrugation case, an additional interference
effect from the interplay between the couplings in the AA and AB region can lead to
another form of localization, which is not of primary interest in the present work.
This interference driven localization effect is discussed in
App.~\ref{app:corrugation}.}, we now discuss how increasing the interlayer coupling $t_\perp$ brings the
three regimes discussed above to life. 


\subsection{Wave function statistics}

The degree to which a wave function, defined for a set of lattice sites $\vec{Q}$, is delocalized is conveniently quantified by the inverse participation ratio (IPR)\cite{RevModPhys.80.1355}, $\sum_{\vec{Q}}|\psi_{\vec{Q}}|^4$. In the  limiting cases of fully delocalized and perfectly localized states, this quantity assumes the values $1/L^2$ and $1$, respectively. More generally, the IPR probes the inverse square of the localization length. More precisely, we define  the localization length $\xi_{\parallel} (\epsilon)$ along a ring of given energy $\epsilon$,
by the inverse participation ratio projected onto the ring as
\begin{align}
\xi_{\parallel} (\epsilon)= 1/ \overline{\Bigg(\sum_{S_l} \Big(\sum_{(\vec{Q} +\vec{K}^{\alpha})\in S_l (\epsilon)} \big |A_n^\alpha (\vec{k}-\vec{Q}) \big|^2 \Big)^2 \Bigg)} \,.
\end{align}
Here, the first sum runs over a set of momentum vectors $S_\ell$ given by $\vec{Q} +  \vec{K}^{\alpha}$ that fulfill the condition 
\[
  \textrm{Round} \left [ \epsilon / (v_F q_m) \left ( \textrm{Arg} (\vec{Q} +  \vec{K}^{\alpha})  + \pi \right)\right] = \ell \quad {\rm  for} \quad \ell \in \mathbb{Z} \,,
\]
where $q_m = |\vec{K}^U -\vec{K}^L |  $ is a momentum difference between momenta in the upper layer, $\vec{K}^{U}$, and lower layer, $\vec{K}^{L}$.
The overbar represents the averaging over momenta $\vec{k}$ in the first \moire Brillouin zone and band index $n$ with eigenenergy
$\epsilon_n (\vec{k})$ in the vicinity of $\epsilon$.

\begin{figure}[t]
	\centering
	\includegraphics[width=\columnwidth]{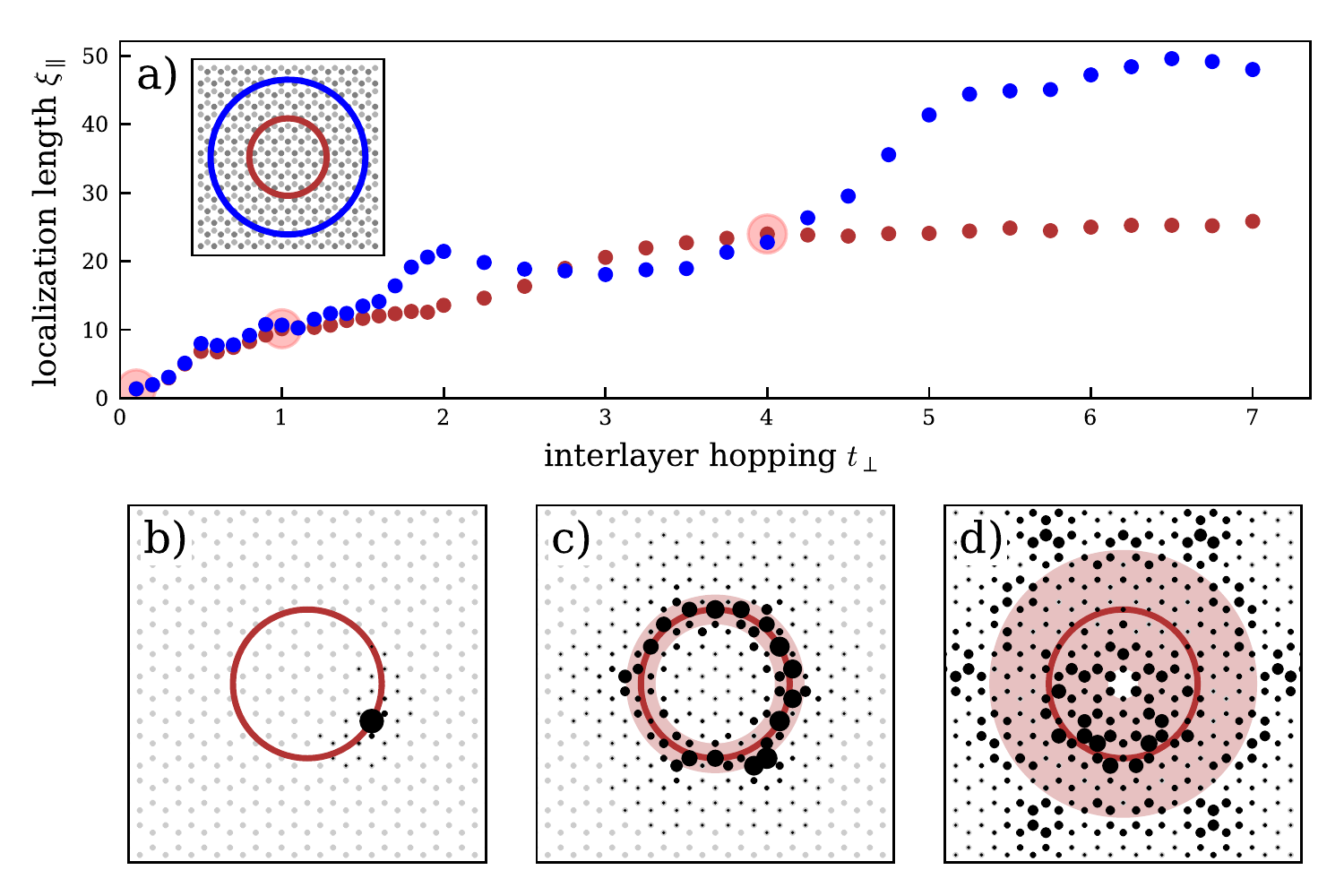}
	\caption{{\bf Momentum space localization.}
	(a) Inset: Momentum space one-dimensional rings in which wave functions are confined.
	The rings are embedded into a bipartite \moire lattice indicated by the gray dots
	(the bipartition is inherited from the $K$ points of the upper and lower graphene layers). Each ring has radius $r = \epsilon / (v_F q_m)$ (energy of the original graphene bands)
	with width $2 t_{\perp} / (v_F q_m)$.
	Main panel: Localization length $\xi_{\parallel}$ in $t_{\perp} / (v_F q_m)$  (blue dots with $r = r_1=10$ and red dots with $r = r_2=5$).
	(b-d) Representative wave functions (black dots) of energy corresponding to the radius  $r_2$, but different $t_{\perp} / (v_F q_m)$:
	(b) $t_{\perp} / (v_F q_m) = 0.1$, (c) $1$, and (d) $4$.}
	\label{Fig:MomentumSpaceLocalization}
\end{figure}

This localization length $\xi_{\parallel}$ is plotted as a function of the interlayer
hopping $t_{\perp}$ in the top panel (a) of Fig.~\ref{Fig:MomentumSpaceLocalization}
for two different dimensionless ring radii $r \equiv \epsilon/(v_F q_m)$, $r_1 = 10$
(blue dots) and $r_2 = 5$ (red dots). The most obvious structure to notice is an
increase of the localization length followed by the eventual saturation at a plateau. For small $t_{\perp}$, the IPR does
not show significant energy dependence, and the two curves approximately coalesce. We interpret this observation as localization driven by the local {\em incommensurability} of the
lattice structure, cf. Fig.~\ref{Fig:Incommensuratelocalization} in
App.~\ref{app:incomm} for an illustration. Roughly, this means that the extension of
wave functions depends on the geometric orientation of the locally straight Fermi surface
relative to the momentum lattice.  With  increasing $t_\perp$, the wave function
explores larger regions of the Fermi surface. For the system of smaller energy, the
circular geometry of the latter becomes visible at values of the localization length,
where the larger surface still looks approximately straight. This is the reason for
the deviations between the two curves at some intermediate coupling strength.
Eventually saturation of the localization length is expected. Contrary to what one might expect, we observe saturation at a value somewhat different from the ring circumference. The origin of this deviation will be discussed below. We have no convincing explanation for the growing tendency for non-monotonous behavior of the localization length with increasing energy. 

It is illuminating to relate this discussion to the structure of actual wave functions. 
At small interlayer coupling $t_{\perp} \approx 0.1 v_F q_m$,  wave functions are fully localized at single (but arbitrary)
points in momentum space, as shown in panel b) of Fig.~\ref{Fig:MomentumSpaceLocalization}. For intermediate interlayer coupling  $t_{\perp} \sim v_F q_m$, they start to spread out along the ring and become mutually correlated, panel c).  However, a still inhomogeneous distribution of wave function weight implies a characteristic band velocity of $\mathcal{O}(\epsilon/k_F)$ in line with the discussion of the intermediate regime II in the previous section. 
For even larger interlayer couplings $t_{\perp} >  v_F q_m$, the wave functions begin to spread out into the two-dimensional momentum space, cf. panel d). (This excursion into the second dimension may explain the above mentioned numerical discrepancy between the saturation value of $\xi_{\parallel}$ and $2 \pi r$.) At the same time we observe regular features, in the shown example an approximate reflection symmetry at the vertical axis. 
These structures herald the increasing importance of lattice symmetries and the entrance into regime III. 


\subsection{Spectral statistics}
\label{sec:SpectralStatistics}

\begin{figure}[t]
	\centering
	\includegraphics[width=\columnwidth]{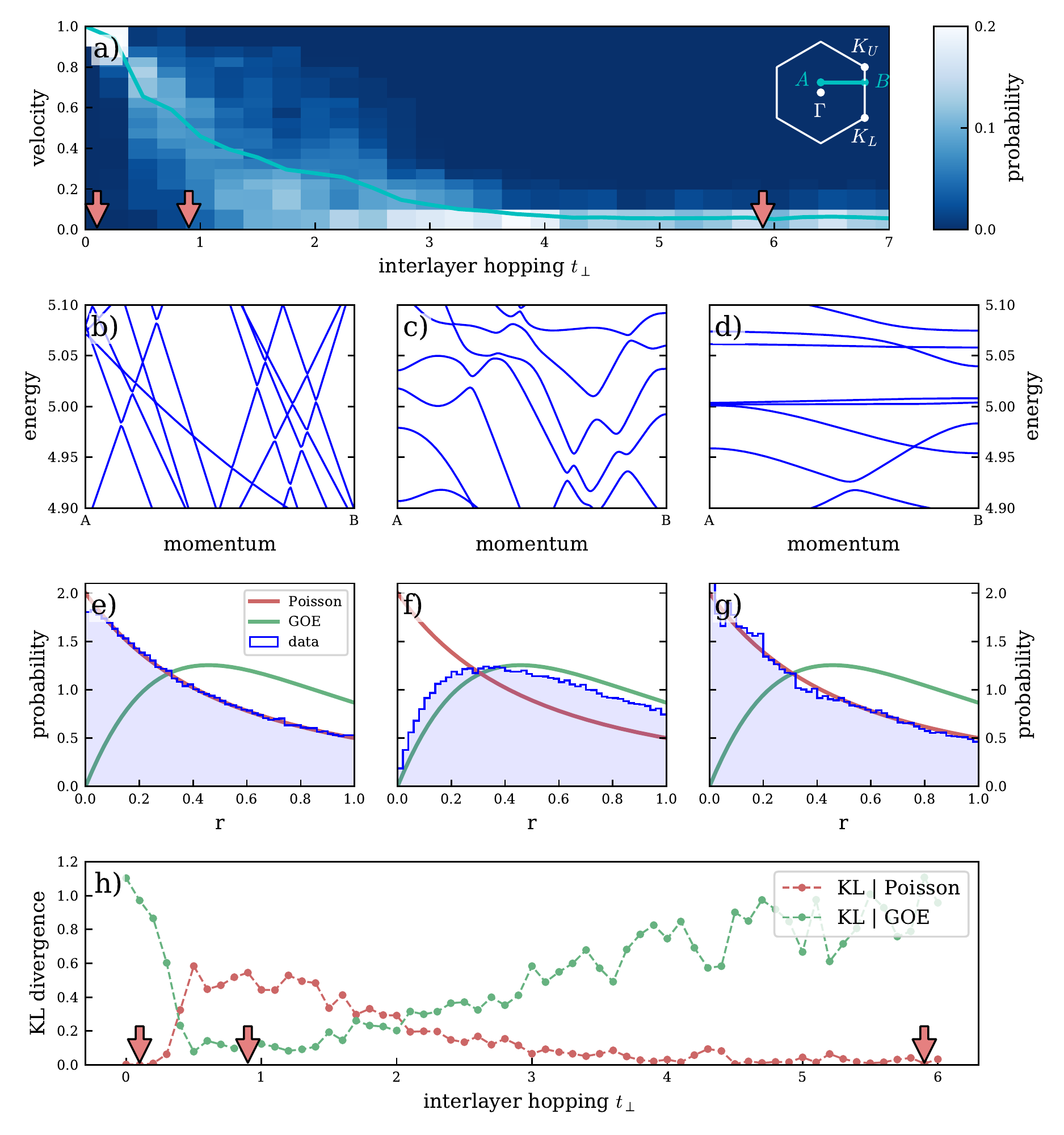}
	\caption{{\bf Spectral statistics.}
			(a) Velocity distribution and the average velocity (cyan trace) as a function of the (dimensionless) interlayer coupling
				$t_{\perp} / (v_F q_m)$.
			(b-d) band structures around $r = 5$ on a momentum cut $A-B$ in the first \moire Brillouin zone (inset)
				for different values of the interlayer coupling (indicated by the arrows in the top panel) with
			(b) $t_{\perp} / (v_F q_m)= 0.1$, (c) $1$, and  (d) $6$.
			(e-g) Level statistics for the parameters corresponding to panels (b-d),
				compared to Poisson statistics (green) and Wigner-Dyson/Gaussian orthogonal ensemble (GOE) distribution (red).
			(h) Normalized Kullback-Leibler (KL) divergences \eqref{eq:KL} calculated 
				for the level statistics.
				The KL divergences are normalized such that 
				$D_{\text{KL}} \left( P_{\text{Wigner}}  || P_{\text{Poisson}} \right)\!=\!1$ and vice versa.}
	\label{Fig:SpectralStatistics}
\end{figure}

To further characterize the different regimes,
we now turn to spectral statistics as a second diagnostic tool.
Our results for the spectral correlations characterizing the system are summarized in Fig.~\ref{Fig:SpectralStatistics}.

The top paned (a) provides an overview of how the band velocities are distributed in dependence of  of the interlayer
coupling $t_\perp$.  Starting from the Fermi velocity (set to 1 in this plot), the average velocity slowly decreases until it asymptotically hovers around a small finite value of $v_{\rm avg} \approx 0.05$ for a numerical value of $t_\perp \gtrsim 4$.
While this evolution of the velocity distribution in itself might not point at three different regimes, such a distinction becomes 
apparent when looking at typical band structures for increasing values of the interlayer coupling as depicted in panels (b-d)
in the middle row of Fig.~\ref{Fig:SpectralStatistics}.

Panel~(b) on the left shows the scenario of `uncooked spaghetti', regime I, with
different energy levels criss-crossing one another at small interlayer coupling. Here
the average band velocities are of the order of $v_F$ and level repulsion is
exponentially small. That this is a momentum space localized regime becomes evident
in the {\em level statistics} of the spectrum. To this end, we calculate the
distribution of ratios of adjacent level spacings $r_n = \Delta E_n / \Delta E_{n+1}$
\cite{PhysRevB.75.155111} (in order to avoid level unfolding). The resulting
distribution is then compared to what is expected for these ratios in Poisson or
Wigner-Dyson statistics (Gaussian orthogonal ensemble, GOE), i.e. to what is expected for a localized versus delocalized
phase, respectively. As demonstrated in the lower row of panels, the data almost perfectly follows the Poisson distribution of a localized
phase.

Further increasing the interlayer coupling one enters the regime II of `semi-cooked spaghetti' with a significant amount of level repulsion
and still steep average dispersion. This is the regime which we identified as the momentum-space localized regime 
above where the wavefunction is
restricted to a ring in momentum space, i.e. delocalized in one dimension but localized with regard to the perpendicular (radial) direction.
In terms of level statistics, this delocalization becomes evident in the observation of Wigner-Dyson statistics, more precisely a Gaussian orthogonal ensemble (GOE) distribution of energy level ratios as shown in panel (f) in the lower row of Fig.~\ref{Fig:SpectralStatistics}.

Finally, for yet stronger coupling we enter the `cooked spaghetti' regime
III, with weakly dispersive bands. Naked eye inspection of the dispersion reveals the
presence of level repulsion, but also level crossings. We tentatively interpret this observation in terms of the almost decoupled symmetry multiplets discussed in the end of section \ref{sec:MomentumSpaceLocalization}. Notice that the weak dispersiveness ($k$-dependence) reflects the relatively weaker influence of the effective disorder: it is no longer strong enough to localize, giving the wave functions a symmetric a roughly isotropic distribution along the fermi surface. Under these circumstances, velocity expectations values are small and depend only weakly on the parameter $k$ entering the lattice potential.  The observed Poisson statistics indicates the absence of correlations between different symmetry sectors of the Hilbert space. However, we have not managed to identify subspaces of cleanly realized lattice symmetries, indicating that the Hilbert space decomposition is only approximated.

In more quantitative terms \footnote{ Using the Kullback-Leibler divergence as a
quantitative measure for describing the level statistics has been introduced in two
recent studies~\cite{monteiro_fock_2021,Berke2020} where it has also been shown
\cite{monteiro_fock_2021} that visual inspections of the level statistics may trick
one into false conclusions. }, the identity of the different regimes can be resolved by monitoring the proximity of the spectral distribution  to either Poisson or Wigner-Dyson
statistics via an entropic measure. To this end, we consider the Kullback-Leibler (KL) divergence
\cite{mezard_information}
\begin{equation}
	D_{\rm KL}(P||Q) = \sum_s p_s \log\left(\frac{p_s}{q_s}\right) \,,
	\label{eq:KL}
\end{equation}
as a logarithmic measure for the difference between two distributions $P$ and $Q$. 
Specifically, the lowest panel in Fig.~\ref{Fig:SpectralStatistics}~(h) shows the KL divergences between the observed distribution and the 
Poisson/Wigner-Dyson distribution as a function of the interlayer coupling.
The crossover from regime I to regime II can clearly be seen in the crossover from Poisson to Wigner-Dyson type distributions. 
However, we also notice that neither of the limiting distributions is generically realized in pure form. 
This is particularly true for regime III, where we observe strong fluctuations of the KL divergences. 
This sensitivity to small variations of the interlayer coupling reflects the fact that spectral statistics,
in particular chaotic regimes following Wigner-Dyson statistics, are particularly sensitive to the presence of discrete symmetries.
The formation of decoupled symmetry multiplets in regime III therefore leads to the return of Poisson statistics for large interlayer couplings.

\subsection{Beyond statistics} 
\label{sec:beyond_statistics}

We already mentioned that the spectra of \moire systems feature structures outside the statistical approach. Indicated by van Hove singularities, visible as spikes in the spectral density shown in Fig.~\ref{Fig:TBG}, these include different types of anomalously flat bands. 

We first note that close to some of their minima the unperturbed bands are approximately parabolic, $\sim p^2/(2 M)$ (in symbolic notation neglecting the two-dimensionality of the problem). The coupling  between the layers adds an effective periodic potential $\sim\!t_\perp\!\cos(G_{m} r)$, defining an  quantum mechanical washboard Hamiltonian.  Referring for a more detailed discussion to Appendix~\ref{app:ladders}, the potential term for relevant model parameters is effectively strong. Under these circumstances we obtain a spectrum comprising bands exponentially small in the ratio of $t_\perp$ and the recoil energy $E_R=G_{m}^2/(2 M)$, $t_\perp /E_R \gg 1$, centered around the equidistant levels of the parabolic potential minima. Examples of such bands are visible in panel h) of Fig.~\ref{Fig:Waterfalls}. 

Then there is of course the celebrated flat band of \emph{magic angle} graphene. In the reading of this paper, this band is the result of an `magically' high level of fine tuning resulting in a band with exceptionally low, but not vanishing $k$-dependence. Besides fine tuning, a factor supporting the flatness of this band is its positioning at zero energy. The average particle-hole symmetry of the spectrum visible in Fig.~\ref{Fig:TBG} implies a tendency for `locking' at this value. At the same time, this band sits at a `Fermi circle' of vanishing  radius, which makes momentum space localization a non-issue. At any rate, the universal principles addressed in this paper have nothing of relevance to say on the engineering of this type  of magically flat bands. 


\section{Discussion} 
\label{sec:discussion}

In this paper, we applied statistical concepts to the description of \moire band
structures. Our starting point was the observation that the incommensurability of the
inter-layer coupling relative to the uncoupled system's dispersion is a source
of effective disorder. On this basis, we defined three regimes governed by different principles: Regime I where weak
interlayer coupling leads to momentum space localization and approximately linearly
dispersive statistically independent energy bands, regime II at intermediate coupling
where states become mutually correlated and partially delocalized  leading to a
nonlinearly dispersive spectrum, and regime III where strong interlayer coupling
makes the symmetries of the \moire lattice a relevant feature and the energy bands
become nearly, but not perfectly flat.

Embedded in the tangle of the statistical spectrum there exist various non-generic features, notably bands of exceptional flatness localized near band extrema. We also interpreted the celebrated magic angle flat band as a genuinely anomalous structure which owes its existence to multi-parameter fine tuning rather than to a universal principle.

On the basis of the above discussion the best bet for encountering an accumulation of exceptionally flat bands is regime III. However, that regime requires $t_\perp \gtrsim \epsilon_F$, i.e. huge interlayer coupling, or Fermi energies scaled to the vicinity of an effective Dirac point (as is the case for the  magic angle flat bands.) 

Some open questions that merit further exploration include the role of screening and its band structure effects. 
One might also entertain the question whether one can engineer longer-range hopping in momentum space
in order to drive the system into the delocalized, chaotic regime already for small interlayer coupling and 
thereby effectively flatten all bands. 
This could possibly be achieved by a manipulation and rearrangement of atoms (while keeping the underlying \moire periodicity unchanged) to  generate a short-range potential in real space, which in turn induces a long-range hopping in momentum space. 


\begin{acknowledgments}
\noindent
{\em Acknowledgments.---}
We thank C. Berke and S. Ilani for insightful discussions.
We acknowledge partial support from the Deutsche Forschungsgemeinschaft (DFG)
-- project grants 277101999 and 277146847 -- through SFB 1238 (project C02) and
within the CRC network TR 183 (projects A01, A03, and A04).
The numerical simulations were performed on the CHEOPS cluster at RRZK Cologne and
the JUWELS cluster at the Forschungszentrum J\"ulich.
\end{acknowledgments}


\bibliography{moirematerials}

\appendix


\section{Magic-angle graphene}
\label{sec:TwistedBilayerGraphene}
We here define the model of twisted bilayer graphene which serves as a basis for the
numerical studies of this paper. Graphene has a lattice constant $a$ of the order of
angstrom and a total bandwidth $D\!\sim\!\mathcal{O}(10\,{\rm
eV})$~\cite{RevModPhys.81.109,PhysRevB.87.205404}. At the magic angle of
$1.1^{\circ}$, the stacked material has a moir\'e lattice constant $a_m\approx
50a$~\cite{bistritzer2011moire,PhysRevB.86.125413}. A moir\'e potential $V$ is caused
by the interlayer coupling $t_\perp\!\sim\!\mathcal{O}(0.1{\rm eV})\!\sim\! D/L$,
which is subject to spatial variations on the large scale of the moir\'e unit cell.
At such small angles, incommensuration effects coming from $L\,G_m \neq G$ are
suppressed by a factor $\left(t_\perp/D \right)^L$, i.e. no quasicrystal physics
occurs. 
The variation of the interlayer coupling may be amplified through the corrugation of the layers in the out-of-plane direction, e.g., the interlayer distance is slightly different in the AA-stacked regions and in the AB-stacked regions, see App.~\ref{app:realspace}.
We take this into account in our modeling, below.
Note that the bandwidths of the much discussed flat bands at the Fermi level in magic-angle graphene are reported in a range of 20 to 40\,meV~\cite{choi2019electronic,jiang2019charge}.
This is significantly larger than theoretical predictions of $\lesssim 10\,{\rm meV}$~\cite{koshino2018maximally}, including corrugation but not in-plane lattice relaxation effects.
Putting the experimentally reported bandwidth into context with $G_m$, the typical velocities $v$ in these bands are only moderately suppressed when compared to the typical velocity $v_F$ of single-layer graphene, i.e. $v \sim 0.1, \ldots, 0.2 \cdot v_F$.

\section{Real-space lattice model} \label{app:realspace}

To perform band structure calculations for twisted bilayer graphene, we start from a real-space lattice model by considering a honeycomb lattice with lattice vectors
\begin{equation}\label{eq:latvec}
\vec{a}_1 = a\left(\frac{\sqrt{3}}{2},-\frac{1}{2}\right), \quad
\vec{a}_2 = a\left(\frac{\sqrt{3}}{2},+\frac{1}{2}\right)
\end{equation}
and lattice constant $a$.

Twisting two graphene layers with respect to each other by a finite angle $\theta$ gives rise to a large-scale interference pattern -- the moir\'e pattern.
At commensurate twist angles, moir\'e \emph{unit cells} are formed which are strictly periodic. 
These commensurate twist angles are achieved by pairs of integer numbers $m$ and $n$ which define the twisting of the Bravais lattice site $R_u = m \vec{a}_1 + n \vec{a}_2$ in the upper layer on top of site $R_l = n \vec{a}_1 + m \vec{a}_2$. The twist angle  $\theta_{m,n}$ for a given pair $(m,n)$ is then defined as
\begin{equation}\label{eq:anglemn}
	\cos\left(\theta_{m,n}\right)  =  \frac{1}{2}\frac{m^2 + n^2 + 4mn}{m^2 + n^2 + mn}\,.
\end{equation}
The corresponding moir\'e Bravais lattice vectors are 
\begin{equation}
\vec{a}_{1,M} = m \vec{a}_1 + n \vec{a}_2, \quad
\vec{a}_{2,M} = R(60^\circ) \vec{a}_{1,M} 
\end{equation}
We further use a generic tight-binding Hamiltonian with all-to-all hopping amplitudes $t(\vec{r}_{i,j})$, reading
\begin{equation}
	\mathcal{H} = \sum_{i,j} t(\vec{r}_{i,j}) c_j^\dagger c_i^{\phantom \dagger}\,.
\end{equation}
The distance-dependent hopping amplitudes are chosen to be in the Slater-Koster form~\cite{PhysRev.94.1498,de2009localization,PhysRevB.87.205404}, i.e.
\begin{align}
	&t(\vec{r}) = V_{pp\pi}(r)\!\left(\!1\! -\!\left(\!\frac{\vec{r}\!\cdot\!\vec{e}_z}{r}\right)^2 \right) + V_{pp\sigma}(r)\!\left(\!\frac{\vec{r}\!\cdot\!\vec{e}_z}{r}\!\right)^2,\\
	&V_{pp\pi}(r) = V_{pp\pi}^0 e^{-(r - a_0)/\delta_0}\,,\\
	&V_{pp\sigma}(r) = V_{pp\sigma}^0 e^{-(r - d_0)/\delta_0}\,.
\end{align}
Here, we have introduced the intralayer nearest-neighbor distance $a_0 = a/\sqrt{3}$, the mean interlayer distance $d_0$ (see next paragraph), the decay distance of orbital overlap $\delta_0$, as well as the two overlap integrals $V_{pp\pi}^0$ and $V_{pp\sigma}^0$. 
In our calculations we base our parameters on the experimentally determined values of $a_0 = 0.142\;\text{nm}$, $\delta_0 = 0.319\;\text{nm}$ as well as $V_{pp\pi}^0 = -2.7\;\text{eV}$ and $V_{pp\sigma}^0 = 0.48\;\text{eV}$.

For the modelling of the real-space structure of twisted bilayer graphene in a setting close to experiments, we further take into account corrugation effects which buckle the lattice on scales of the moir\'e cell, i.e. we introduce a periodic variation of the interlayer distance. 
This effect stems from the interactions between the atoms of the two layers which are stacked either in an AA or AB fashion, depending on the relative position within the moir\'e cell. 
The layer distance therefore varies between $d_{\mathrm{AA}}$ and $d_{\mathrm{AB}}$ with the periodicity of the moir\'e cell as
\begin{align}
	d(\vec{R}) & = d_0 + 2 d_1 \sum_{i=1}^3 \cos\left(2\pi\frac{\vec{R}\cdot\vec{C}_i}{|\vec{C}_i|^2}\right)\,.
\end{align}
Here, $d_0$ and $d_1$ are based on the AA and AB distances as
\begin{align}
	d_0 &= \frac{1}{3} (d_{\mathrm{AA}} +  2 d_{\mathrm{AB}})\,,\\
	d_1 &= \frac{1}{9} (d_{\mathrm{AA}} - d_{\mathrm{AB}})\,.
\end{align}
The corrugation spanning vectors $C_i$ which are enclosing an angle of $60^\circ$ span the moir\'e pattern and read
\begin{align}
	\vec{C}_1 &= \frac{1}{2} (\vec{a}_{1,M} + \vec{a}_{2,M})\,, \\
	\vec{C}_2 &= R(60^\circ) \vec{C}_1\,,\\
	\vec{C}_3 &= R(120^\circ) \vec{C}_1\,.
\end{align}
The AA and AB distances have been experimentally determined to be $d_{\mathrm{AA}} = 0.360\;\text{nm}$ and $d_{\mathrm{AB}} = 0.335\;\text{nm}$~\cite{lee2008growth,PhysRevB.90.155451,koshino2018maximally}.

When referring to specific combinations of model parameters, we always use relative values with respect to the experimental parameters, e.g. a corrugation of $0$ refers to a layer distance of the AB regions everywhere in the lattice. 
In total, we modify the layer distance, the strength of corrugation, as well as the orbital overlap.
To that end, we introduce the interlayer coupling parameter $V$ which denotes the interlayer Slater-Koster parameter $V_{pp\sigma}^0$ in units of its experimental value.
Further, we parametrize the interlayer distance $D_l$ in units of the equilibrium distance as well as the level of corrugation by a parameter $C$ which denotes the corrugated part of the interlayer distance $d_{AA} - d_{AB}$ in units of the experimentally determined value.

\section{One-dimensional moir{\'e} system}
\label{App:1D-Moire}

In this section, we study a one-dimensional moir{\'e} system to explore the effective range of the coupling between lattice sites in momentum space.
We show that this coupling decays exponentially on the scale $\Delta p \gtrsim G_m$, where $G_m$ is the shortest \moire reciprocal lattice vector.

 \begin{figure}[b]
	\centering
	\includegraphics[width=\columnwidth]{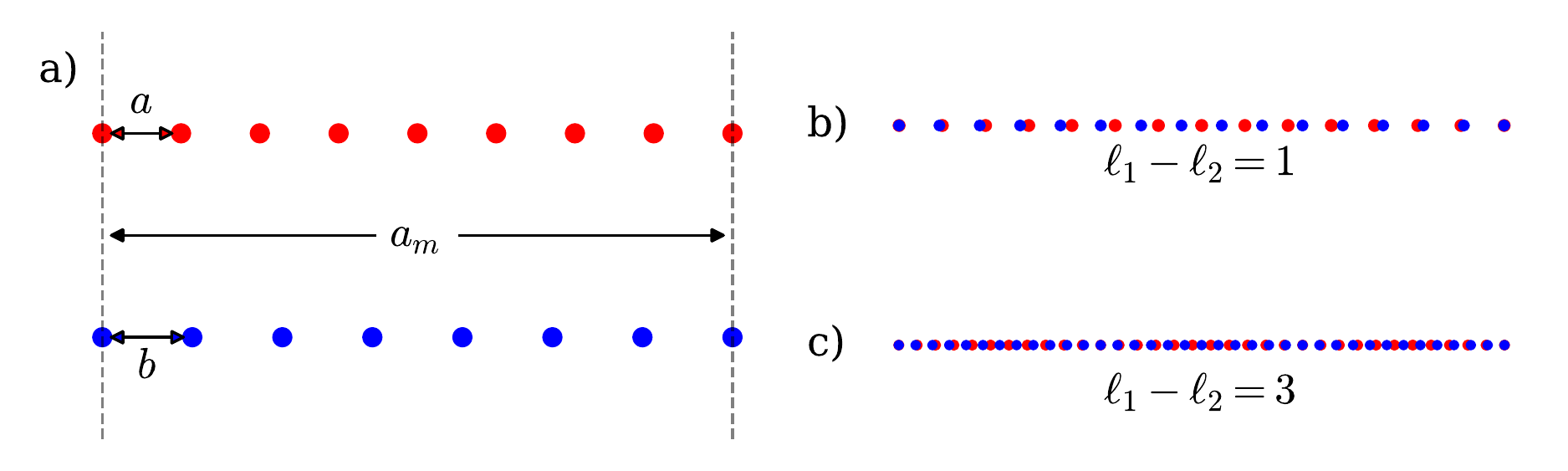}
	\caption{{\bf One-dimensional \moire system.} (a) Two layers with different lattice constant $a$ and $b$. $|a-b|$ is assumed to be much smaller than $a$ and $b$, or equivalently $a, b$ are much smaller than the size of \moire unit cell, $a_m$. The lattice sites in the upper (lower) layer are denoted as red (blue) dots. (b-c) commensurate lattices with the condition $\ell_1 a = \ell_2 b$ with integers $\ell_1, \ell_2$. b) $\ell_1 = 15$ and $\ell_2 = 14$. c) $\ell_1 = 36$ and $\ell_2 = 33$. There is a 3-fold approximate symmetry within the exact periodicity. After washing out the fine structures of the lattices by a bootstrap (e.g., disorder), the periodicity of the approximate symmetry governs features of the coupling of lattice sites in the momentum space.}
	\label{Fig:1dmoire}
\end{figure}

To that end, we consider a one-dimensional moir{\'e} system consisting of two chains  
with different lattice constants $a$ and $b$.
Each layer feels a `substrate potential' induced by
the other layer, cf.~Fig.~\ref{Fig:1dmoire}~a).
To obtain a large-scale moir\'e interference pattern,
$|a - b|$ is assumed to be much smaller than $a$ and $b$.

Here, we first focus on the incommensurate situation where $a/b$ cannot be written in the form $p/q$ with $p, q\in \mathbb{Z}$. The length of the \moire unit cell is given by $a_m = ab/|a-b|$ and the length of the unit of the \moire reciprocal lattice vector $G_m = 2 \pi / a_m$. 
We also comment on the commensurate situation, below.

\begin{figure*}[t]
	\centering
	\includegraphics[width=.75\linewidth]{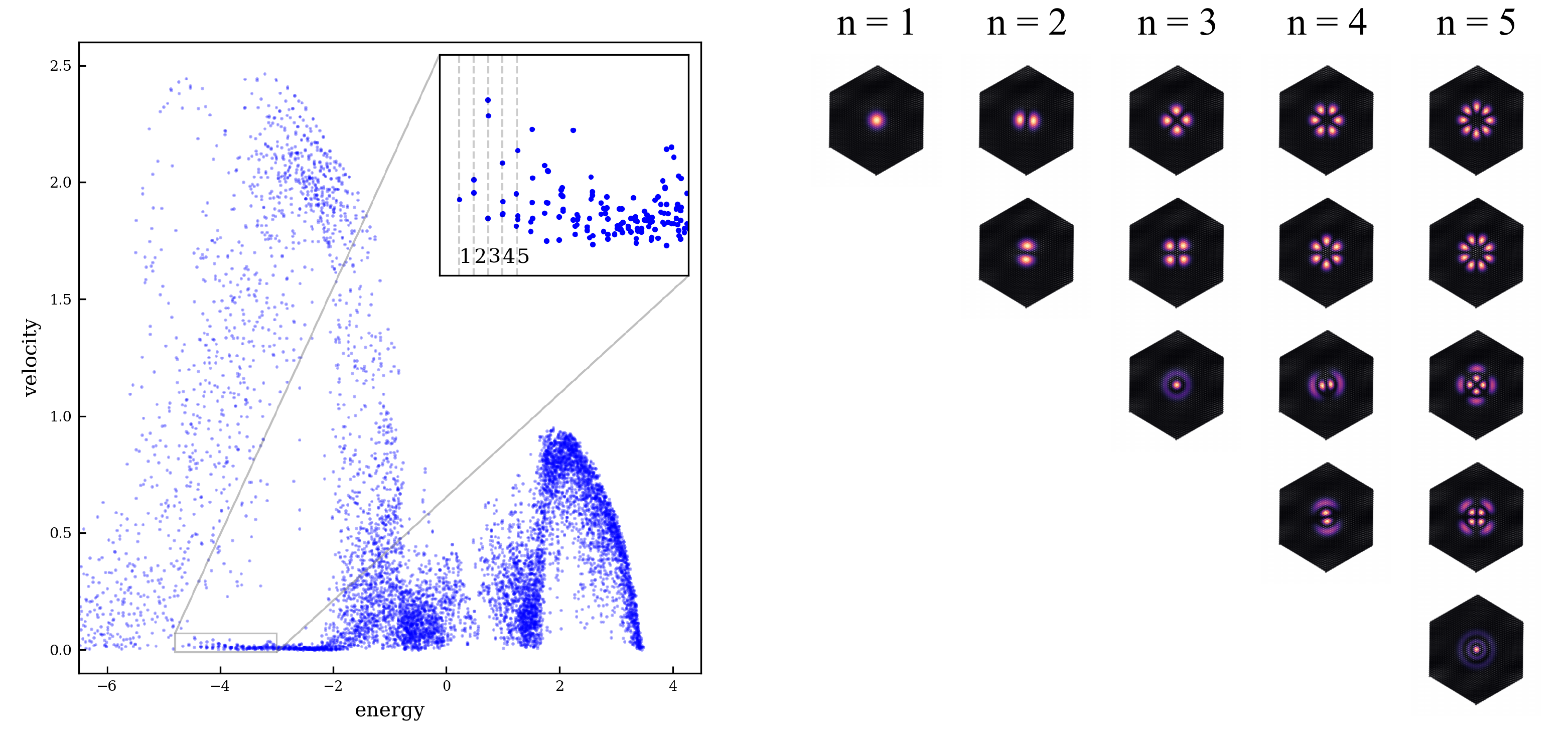}
	\caption{{\bf Formation of ladder states} near band minimia.
			The `waterfall' plot (akin to Fig.~\ref{Fig:Waterfalls} in the main text) 
			on the left highlights the formation, at finite interlayer coupling, of a sequence of 
			equally-spaced flat bands emerging near the minimum of the graphene band of the uncoupled system.			 
			The cascade of plots on the right illustrates the wavefunctions for the first 5 energy levels 
			extracted from exact diagonalization of our real-space TBG model.
			The parameters are identical to those in Fig.~\ref{Fig:Waterfalls} panels d) and h), i.e. strong interlayer coupling $V=2$ and strong corrugation $C=5$.
	}
	\label{Fig:FlatBandLadders}
\end{figure*}

The potential $V$ in the upper chain as induced by the lower chain reads
\begin{align}
V(R^{U}) = \sum_{R^{L}} U (R^{U} -R^{L} )\,,
\end{align}
where $U (R^{U} -R^{L} )$ is the microscopic interlayer coupling which only depends on the distance $|R^{U} -R^{L}|$.
Through the potential $V$ an electron with initial momentum $p$ can be scattered into a state with momentum $p'$. This process is described by the matrix element
\begin{align} \label{hopping1d}
\langle p' | V | p \rangle & =\frac{1}{N} \sum_{R^U} V(R^U) e^{i (p- p') R^U}  \nonumber \\
				    & =\frac{1}{N} \sum_{R^U, R^{L}} U(R^{U} -R^{L}) e^{i (p - p') R^U}\,,
\end{align}
where $N$ is the number of sites in the upper chain. To study the behavior of this matrix element for long wavelengths, we introduce a smoothing function $f(x)$~\cite{PhysRev.94.1498}.
The smoothing property of $f$ allows us to treat the moir\'e potential as a continuous and smooth modulation and estimate the induced hopping range.
To realize the smoothing, $f(x)$ is chosen to decay on a scale  much larger than the atomic scale $a$, but much smaller than the moir\'e scale $a_m$ and it is normalized as $\int_{a_m} dx f (x) = a_m$.
Inserting this normalization condition into Eq.~\eqref{hopping1d} yields
\begin{align}
\langle p' | V | p \rangle &=\!\frac{1}{Na_m}\!\int_x\!\sum_{R^U\!,R^{L}}\!f(x\!-\!R^U)U(R^{U}\!-\!R^{L})e^{i (p - p') R^U} 
\nonumber \\ 
&\sim  \frac{1}{N a_m}\!\sum_{R^U\!, R^{L}}\!\int_x e^{i (p - p') x} f(x\!-\!R^U)U(R^{U}\!-\!R^{L}) \nonumber \\  
& = f_{p-p'}\sum_{G^U, G^{L}, q} U(q) \delta_{p-p'+q, G^U} \delta_{q, G^L}.\label{eq:kron}
\end{align}
Here, we used $e^{i (p - p') R^U}\sim e^{i (p - p') x}$ when going from the first to the second line as the smoothing function acts as a delta function on the moir{\'e} scale and we introduced the reciprocal lattice vectors of the upper/lower chains $G^{U/L}$.
The two Kronecker deltas in the last line of Eq.~\eqref{eq:kron} lead to the relations
\begin{align}
\Delta p = p -p' &= G^U- G^L = 2 \pi \left(\frac{m_1}{a} - \frac{m_2}{b}\right) \equiv Q, \nonumber \\
q &= G^L = 2 \pi \frac{m_2}{b}\,,\label{eq:decay}
\end{align}
with $m_1, m_2 \in \mathbb{Z}$ and the \moire reciprocal lattice vectors~$Q$.
To obtain a non-negligible contribution to $\langle p' | V | p \rangle$, the smoothing function $f (\Delta p)$ needs to be sizeable, which is the case for small momentum transfer $\Delta p$ smaller than the decay length ($\ll 1/a$ or $1/b$) of $f(\Delta p)$.
This is realized for the case $m_1=m_2$, which we discuss in the following.

Since $U (x)$ decays on the atomic scale $a$, the potential $U(q)$ decays on the scale $\sim 1/a$. 
Therefore, the dominant contribution comes from the $q = 0$ part, but it gives rise to the condition $p = p'$, and thus results in a trivial energy shift by a constant term.
The lowest non-trivial contribution comes from the $m_1 =m_2=\pm 1$.
According to Eqs.~\eqref{eq:decay}, we then obtain $\Delta p = \pm 2\pi (1/a-1/b)$ and $q = \pm  2\pi /b$ leading to the matrix element $\langle p' | V | p \rangle = U (2\pi/a)$.
More generally, choosing $m = m_1= m_2$, i.e. $\Delta p  = 2 \pi m (1/a-1/b)$ and $q=2\pi m/b$, yields the matrix element
\begin{align}
\langle p' | V | p \rangle = U \left (\frac{2\pi m}{b} \right)\,.
\end{align}
Hence, together with the assumption that the microscopic interlayer coupling $U$ decays exponentially on the atomic scale $a$, the coupling of the lattice sites in momentum space decays exponentially over scale $\Delta p \gtrsim G_m$.

It is instructive to see how this argument can be generalized to the commensurate case, where $\ell_1 a = \ell_2 b$ with integers $\ell_1$ and $\ell_2$. 
When $|\ell_1 - \ell_2 | = 1$, cf.~Fig.~\ref{Fig:1dmoire}~b), the lowest non-trivial condition $p-p' = 2\pi (1/a-1/b)= \pm 2\pi/(\ell_1 a)=\pm G_m $ is fulfilled; 
the range of the coupling of the lattice sites in momentum space decays over the scale $G_m$. 
Even for $|\ell_1 - \ell_2 | \neq 1$, we always find the approximate symmetry that satifies $\ell'_1 a \simeq \ell'_2 b$ with $|\ell'_1 - \ell'_2 | = 1$, cf.~Fig.~\ref{Fig:1dmoire}~c), only if $|a-b| \ll a, b$. This periodicity of the approximate symmetry governs features of the coupling of lattice sites after washing out the fine structures of the lattices by a possible bootstrap, e.g., disorder. Therefore, the statement that the coupling of lattice sites in momentum space decays over the scale $\Delta p \gtrsim G_m$ holds generally for \moire systems, regardless of whether the the lattice configuration is commensurate or incommensurate.

\section{Flat band ladders}
\label{app:ladders}

An exceptional feature of our real-space TBG model is the occurrence of a sequence of equally-spaced flat bands.
This sequence emerges near the minimum of the graphene bands of the uncoupled system, cf.~ the inset in panel~h) of Fig.~\ref{Fig:Waterfalls} in the main text.
The underlying physical mechanism at play here can be readily understood in terms of a harmonic-oscillator level spacing.
To that end, consider electrons near the band minimum of the uncoupled layers tunneling in the periodic moir\'e potential, 
which -- along one chosen direction -- is dominated by the modulation term $\sim\!t_\perp\!\cos(G_{m} r)$~\footnote{We note that the full argument for the two-dimensional case, based on the consideration of both basis vectors of the reciprocal lattice, 
$\vec{G}_{m,1}$ and $\vec{G}_{m,2}$, works analogously.}.
Near a band minimum, the kinetic energy of the electrons can be approximated using a quadratic approximation $\sim\!k^2/(2m)$, where $m$ is given by the curvature at that minimum.
The potential is large as compared to the recoil energy, since the parameters in the presented model imply 
\[
	t_\perp \gtrsim 1/L \gg G_{m}^2/(2m) \sim 1/L^2 \,.
\]
This allows us to extract the corresponding level spacing of the harmonic oscillator eigenenergies yielding 
\[
	\sqrt{t_\perp G_{m}^2/m}\sim 1/L^{3/2} \gg 1/L^2 \,.
\]
In consequence, we obtain a sequence of equally spaced bands with an exponentially small bandwidth 
\[
	\Delta W \sim \exp(-\sqrt{t_\perp m}G_{m}) \sim \exp(-\sqrt{N}) \,.
\]
Definite numerical evidence of this scenario is provided in Fig.~\ref{Fig:FlatBandLadders}, where we show the wave-function solutions of the eigenstates in this sequence 
as plotted in the cascade of panels on the right -- a beautiful illustration of two-dimensional harmonic oscillator states.

 \begin{figure*}[th!]
	\centering
	\includegraphics[width=\linewidth]{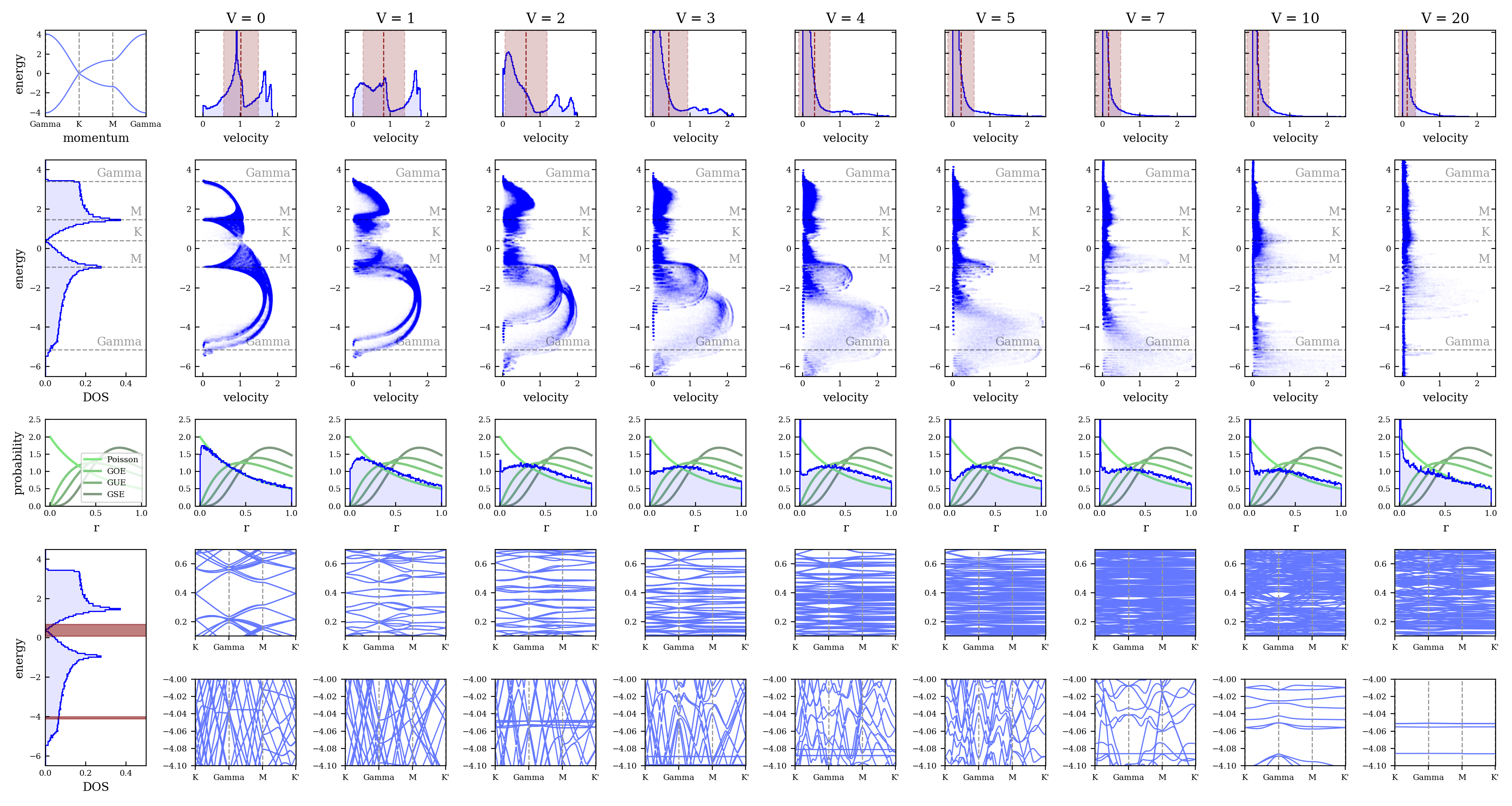}
	\caption{{\bf Spectral statistics for real-space TBG model as a function of interlayer coupling} 
	at an angle corresponding to $m=15, n=12$, cf.~\eqref{eq:anglemn}, i.e. $\theta \approx 2.28^\circ$. We have choosen an interlayer distance of 55\% of the experimental distance and a corrugation that is five times stronger than the experimental value, i.e. $D_l = 0.55$ and $C=5$, cf. Sec.~\ref{app:realspace}.
			Columns are shown from left to right with increasing interlayer coupling parameter $V$ 
			which denotes the interlayer Slater-Koster parameter $V_{pp\sigma}^0$ in units 
			of the experimental value $V_{pp\sigma}^0 = 0.48\;\text{eV}$.
			The upper two rows show velocity statistics whereas the lower three rows illustrate the spectral statistics. 
			The middle row shows histograms of $r$ values which are compared to Poisson or Wigner Dyson statistics 
			indicate localization versus chaotic regimes. The underlying histograms of values for the ratios of adjacent level spacings~$r$ are computed by averaging 
			over several momenta within the first \moire Brillouin zone as well as all energies. 
			The spectral statistics of the real-space model are in overall agreement with what is discussed for the momentum space model in Sec.~\ref{sec:SpectralStatistics} in the main text.}
	\label{Fig:RealSpaceSpectralStatistics}
\end{figure*}

\section{Spectral statistics for TBG}
\label{app:SpectralStatistics}

We complement the analysis of spectral statistics for the continuum model in the main text (Sec.~\ref{sec:SpectralStatistics})
with numerical data for the real-space TBG model here. While such an analysis for the real-space model does not give the
same quantitative clarity as for the continuum model, we observe very similar qualitative behavior as a function of increasing 
interlayer coupling.
A summary is provided in Fig.~\ref{Fig:RealSpaceSpectralStatistics}.
Here, the upper two rows show the velocity statistics  as a function of increasing interlayer coupling (different columns), cf.~Fig.~\ref{Fig:Waterfalls} of the main text.
Therein, we present a histogram of velocities in the first row and energy-resolved `waterfall' plots in the second row. 
The lower three rows show the spectral statistics.
Here, the two rows at the bottom are snapshots of the band structure in the two energy windows as indicated on the very left.
A quantitative analysis of the level spacings in these band structure plots is provided in the middle row.
Here, we present the level-spacing distribution as a histogram that is compared to the Poisson and Wigner-Dyson distributions, which are characteristic for localized and delocalized chaotic regimes, respectively.

Looking at the evolution for increasing interlayer coupling parameter $V$, we observe a similar progression as discussed for the continuum model in the main text, cf.~Sec.~\ref{sec:SpectralStatistics}:
starting from a Poisson-like distribution the level statistics evolves to a broad histogram
as resembling a Wigner-Dyson-like distribution. 
We note, however, that the formation of flat band ladders, cf.~Sec.~\ref{app:ladders}, gives rise to a $\delta$-function like peak at vanishing ratio of adjacent level spacing, i.e. at $r=0$.
This phenomenon somewhat obscures the evolution of the level spacing distribution. 
For large interlayer coupling, the distribution moves back to a monotonously decaying Poisson-like distribution, 
akin to the reentrance behavior discussed for the continuum model in Sec.~\ref{sec:SpectralStatistics}.

\section{Momentum-space continuum model  for TBG}
\label{app:momentum}

In this section, we review the continuum model for TBG as developed in Ref.~\cite{bistritzer2011moire} as valid for small twist angles $\theta$. We employ this model for the calculations in Sec.~\ref{sec:ContinuumModel} as a simple tool to understand the complex band structures of the twisted bilayer graphene based on statistical principles.

The starting point is a monolayer of graphene with primitive lattice vectors as given in Eq.~\eqref{eq:latvec}.
Explicitly, the basis of the honeycomb lattice sites reads $\vec{\tau}_A = 0, \vec{\tau}_B = a\vec{\hat{\vec{x}}}/\sqrt{3}$ and
the primitive reciprocal lattice vectors are $\vec{b}_1 = (2\pi/a)(\vec{\hat{\vec{x}}}/\sqrt{3}+\vec{\hat{\vec{y}}})$ and $\vec{b}_2 = (2\pi/a)(\vec{\hat{\vec{x}}}/\sqrt{3}-\vec{\hat{\vec{y}}})$, where $\vec{\hat {\vec{x}}}$ and $\vec{\hat{\vec{y}}}$ are euclidean unit vectors in the $x$ and $y$ direction, respectively.
The inequivalent $\vec{K}$ and $\vec{K'}$ points  read
\begin{align}
\vec{K} = \frac{4 \pi}{3 a} \vec{\hat{\vec{y}}},\quad \vec{K'} = - \vec{K} =  -\frac{4\pi}{3  a} \vec{\hat{\vec{y}}}.
\end{align}
Accordingly, the equivalent $\vec{K}$ points in the first Brillouin zone are given by translations by primitive reciprocal lattice vectors, i.e. 
$\vec{K}_1 = \vec{K}, \vec{K}_2 = \vec{K} - \vec{b}_1$, and $\vec{K}_3 = \vec{K} - \vec{b}_1 + \vec{b}_2$.

We now consider a twisted graphene bilayer with total relative twist angle $\theta$. For symmetry reasons, the upper layer is twisted by $+ \theta / 2$ and the lower layer is twisted by $- \theta/2$ with respect to a perfectly aligned AA stacking of the bilayer. 
The Bloch wave function with momentum $\vec{k}$ ($\vec{k}'$)  residing on sublattice  $\beta$ ($\beta'$) in the upper (lower) layer is written as
\begin{align}
| \Psi_{\vec{k} \beta}^{U} \rangle = \frac{1}{\sqrt{N_{U}}} \sum_{\vec{R}^{U}} e^{i \vec{k} \cdot \vec{R}^{U}  }
|\vec{R}^{U} + \vec{\tau}_{\beta}^{U} \rangle,\nonumber \\
| \Psi_{\vec{k}' \beta'}^{L} \rangle = \frac{1}{\sqrt{N_{L}}} \sum_{\vec{R}^{L}} e^{i \vec{k}' \cdot \vec{R}^{L}  }
|\vec{R}^{L} + \vec{\tau}_{\beta'}^{L} \rangle.
\end{align}
Here $N_{U (L)}$ is the number of the unit cells in the upper (lower) layer and $\vec{\tau}_{\beta}^{U (L)} \equiv e^{\pm i \theta \sigma_z/2} \vec{\tau}_{\beta} e^{\mp i \theta \sigma_z/2}$ is the rotated basis. 

\begin{figure*}[t!]
	\centering
	\includegraphics[width=\linewidth]{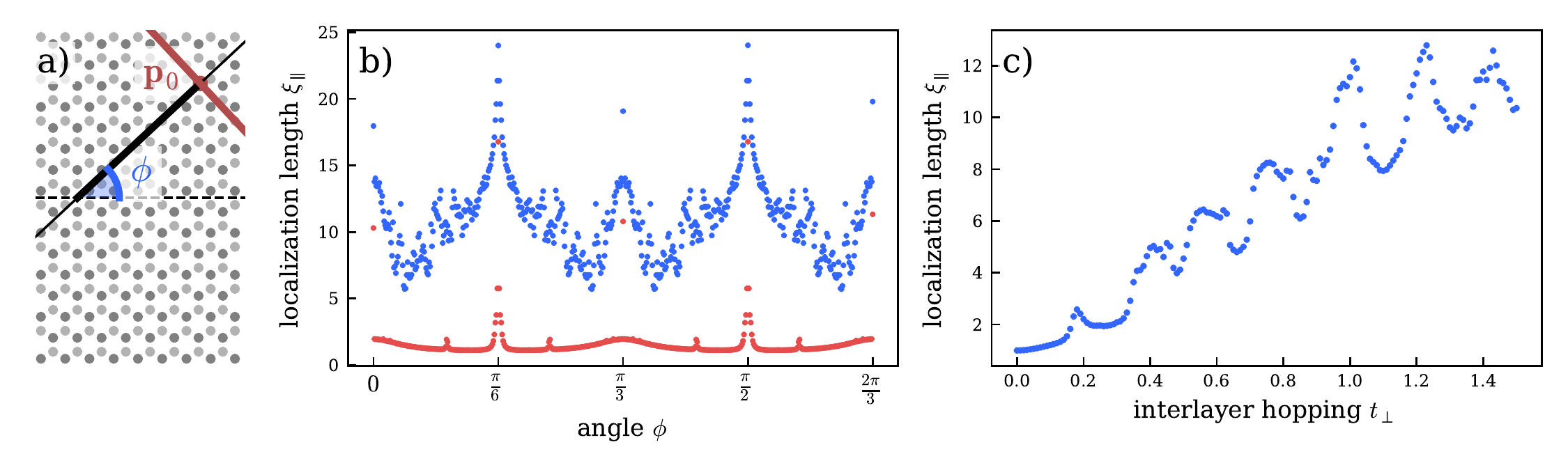}
	\caption{{\bf Localization by incommensurability.} (a) The circle of equal energy crossing momentum $\vec{p}_0 = p_0 (\cos \phi, \sin \phi)$. In the limit of $p_0 \rightarrow \infty$ this can be regarded as a straight line (red line) with angle $(\pi - \phi)$. (b) Localization length $\xi_{\parallel}$ along the line as a function of the angle $\phi$ with pronounced peaks at commensurate angles. Here, the notion '(in)commensurate angle' means that the distances between the lattice points and the line are (in)commensurate. Different colors represent different $t_{\perp}$, i.e. $t_{\perp} = v_F q_m$ (blue dots) and $t_{\perp} = 0.1 v_F q_m$ (red dots). The pronounced peaks of $\xi_{\parallel}$ at commensurate angles show that incommensurability is a dominant source of localization. (c) $\xi_{\parallel}$ as a function of $t_{\perp}$ with fixed $\phi= \pi/12$.}
	\label{Fig:Incommensuratelocalization}
\end{figure*}

The bare graphene (Dirac) Hamiltonian of layer $U (L)$ near the $K^{U (L)}$ point then reads
\begin{align}
H_{U/L}^{K} (\vec{p}) = - \frac{ \sqrt{3} a t}{2} \left (\vec{\sigma}_{\pm \theta/2} \cdot \vec{p} \right),
\end{align}
where $\vec{\sigma}_{\theta} = e^{i \theta \sigma_z/2} \vec{\sigma} e^{-i \theta \sigma_z/2}$.
Given momentum $\vec{p}$ in the first Brillouin zone, the bare graphene Hamiltonian with momentum $\vec{p} - \vec{Q}_n$ can be unfolded into the extended Brillouin zone scheme as
\begin{align} \label{graphenedispersion}
H_{U/L}^{K} (\vec{p} - \vec{Q}_n) = - \frac{ \sqrt{3} a t}{2} \left (\vec{\sigma}_{\pm \theta/2} \cdot \left (\vec{p} - \vec{Q}_n \right) \right ).
\end{align}
Here $\vec{Q}_n$ are the \moire reciprocal lattice vectors, associated with the periodicity of the moir{\'e} unit cell.
 
The interlayer hopping matrix describing a process where an electron with momentum $\vec{k}$ residing on sublattice $\beta$ in the upper layer hops to a state with momentum $\vec{k'}$ on sublattice $\beta'$ in the lower layer reads
%
\begin{align} \label{hopping}
&\langle  \Psi_{\vec{k'} \beta'}^{L} |H_T | \Psi_{\vec{k} \beta}^{U} \rangle
=\!\sum_{\vec{R}^{U}, \vec{R}^{L}}\!\frac{t(\vec{R}^{L}\! +\! \vec{\tau}_{\beta'}^{L}\! -\!\vec{R}^{U}\! -\!\vec{\tau}_{\beta}^{U})}{\sqrt{N_U N_L}}  e^{i \vec{k} \cdot \vec{R}^{U}\!- i \vec{k}' \cdot \vec{R}^{L}} \nonumber \\
&\quad = \sum_{\vec{R}^{U},\vec{R}^{L},\vec{q}} \frac{t_{\vec{q}}^{\beta' \beta}}{N_U N_L} 
e^{i (\vec{q}-\vec{k}') \cdot \vec{R}^{L}\!-i (\vec{q}-\vec{k}) \cdot \vec{R}^{U}\!+i \vec{q} \cdot (\tau_{\beta'}^{L} - \tau_{\beta}^{U})}
\nonumber \\ 
&\quad = \sum_{\vec{G}^L, \vec{G}^U } \sum_{\vec{q}} t_{\vec{q}}^{\beta' \beta} \delta_{\vec{q} - \vec{k}, \vec{G}^U}  \delta_{\vec{q} - \vec{k'}, \vec{G}^L}
e^{i \vec{q} \cdot (\tau_{\beta'}^{L} - \tau_{\beta}^{U})}.
\end{align}
%
Here, $\vec{G}^{U (L)}$ is the reciprocal lattice vector of the upper (lower) graphene layer and we have assumed that $t (\vec{R}^{L} + \vec{\tau}_{\beta'}^{L} -\vec{R}^{U} -\vec{\tau}_{\beta}^{U})$ only depends on the distance $|\vec{R}^{L} + \vec{\tau}_{\beta'}^{L} -\vec{R}^{U} -\vec{\tau}_{\beta}^{U}|$.
The two delta functions in Eq.~\eqref{hopping} lead to the \moire condition
\begin{align}
\vec{k} - \vec{k'} = \vec{G}^{U} - \vec{G}^{L} = \vec{Q}_n.
\end{align}
Under the assumption that $t_{\vec{q}}$ decays rapidly around $|\vec{q}| \sim |\vec{K}|$, $\vec{q}$ can be restricted to momenta around the $K$ points in the first Brillouin zone ($\vec{K}_1$, $\vec{K}_2$ and $\vec{K}_3$); $\vec{q}$ can be decomposed as $\vec{q} = \vec{K}^{L}_{i} + \vec{p}' = \vec{K}^{U}_{i} + \vec{p}$ for $i = 1, 2, 3$ with small momentum $\vec{p}$ and $\vec{p}'$, leading to $\vec{p} - \vec{p}' = \vec{K}_{i}^{L} - \vec{K}_{i}^{U} \equiv \vec{q}_i$.
Within those approximations, Eq.~\eqref{hopping} becomes
\begin{align}
\langle  \Psi_{\vec{p}' \alpha'}^{L} |H_T | \Psi_{\vec{p} \alpha}^{U} \rangle =
\sum_{i = 1, 2, 3} t_{\vec{K}_i}^{\beta' \beta} e^{i \vec{K}_i \cdot (\vec{\tau}_{\beta'}^{L} - \vec{\tau}_{\beta}^{U})} \delta_{\vec{p} - \vec{p}', \vec{q}_i}\,.
\end{align}
We assume that $ t_{\vec{K}_i}^{\beta' \beta}$ are momentum independent ($t_{\vec{K}_i}^{\rm AA}= t_{\vec{K}_i}^{\rm AA} = t^{\rm AA}$ and
$t_{\vec{K}_i}^{\rm AB}= t_{\vec{K}_i}^{\rm AB} = t^{\rm AB}$), and further neglect the small angle dependence of
$\vec{\tau}^{L}_{\alpha}, \vec{\tau}^{U}_{\alpha} \simeq \tau_{\alpha}$. Then, the hopping term becomes a 2$\times$2 matrix, reading
%
\begin{align} \label{hoppingFinal}
\langle  \Psi_{\vec{p}'}^{L} |H_T | \Psi_{\vec{p}}^{U} \rangle &=\! \sum_{i = 1,2,3} \delta_{\vec{p} - \vec{p}', \vec{q}_i } \Big(
t^{\rm AB} \Big[\sigma_x \cos \Big(\frac{2\pi}{3} (i-1)\Big) \nonumber\\
&\quad+ \sigma_y  \sin\Big(\frac{2\pi}{3} (i-1)\Big) \Big] + t^{\rm AA} \mathbb{I}
\Big).
\end{align}
This completes the description of twisted bilayer graphene in the continuum model as employed in Sec.~\ref{sec:ContinuumModel}.

\section{Localization by incommensurability}
\label{app:incomm}

In this section, we address the localization driven by the local incommensurability as mentioned in Sec.~\ref{sec:ContinuumModel}. 
To this end, we consider an approximate Hamiltonian, describing a physical mechanism that takes place in a small fraction of a one-dimensional energy circle of the original graphene dispersion with large radius, cf.~the outer-most ring in the inset of Fig.~\ref{Fig:MomentumSpaceLocalization}~a).

We start by taking a momentum $\vec{p}_0 = p_0 (\cos \phi, \sin \phi)$, cf.~Fig.~\ref{Fig:Incommensuratelocalization}~a), which is large with respect to $\vec{K}$, i.e. the $K$~point of the untilted graphene band.
Then we consider the circle of equal energies with radius $v_F p_0$ passing through momentum $\vec{p}_0$.
In the limit of $p_0 \rightarrow \infty$, the fraction of the circle in the vicinity of momenta $\vec{p}_0$ becomes a straight line with angle $(\pi - \phi)$, cf.~the red line in Fig.~\ref{Fig:Incommensuratelocalization}~a). Then, the curvature of the circle is negligible.

The angle $\phi$ determines the level of incommensurability of the geometrical distances between the line and the sites of the momentum-space lattice.
For example, for $\phi = \pi /3$, the distances are commensurate.
To describe the physics near $\vec{p}_0$, we expand the Hamiltonian in Eq.~\eqref{graphenedispersion} in $\delta \vec{p} = \vec{p} - \vec{p}_0 $ as $H^{K}_{l} (\vec{p}) \simeq H^{K}_{l} (\vec{p}_0) + \delta H^{K}_{l} (\vec{p})$, with $ \delta H^{K}_{l} (\vec{p})$ reading
\begin{align}
\delta H^{K}_{l} (\vec{p}) &=v_F \langle \psi_{\vec{p}_0, l}^{+}|(\delta \vec{p}  + \vec{Q}_n - \vec{K}^{l}) \cdot \vec{\sigma}_{\pm \theta/2}| \psi_{\vec{p}_0, l}^{+} \rangle \nonumber \\
  &=  \frac{v_F}{p_0} (\delta \vec{p}  + \vec{Q}_n - \vec{K}^{l}) \cdot \vec{p}_0\,,
\end{align}
where $l \in \{U,L\}$ and $| \psi_{\vec{p}_0, U/L}^{+} \rangle = \left (\exp \left (- i  (  \phi \mp \frac{\theta}{2}) \right), 1 \right )^T$ is an eigenstate of $H^{K}_{U/L} (\vec{p}_0) $ with positive energy $v_F p_0$.
Likewise, the effective interlayer coupling is given by
\begin{align}
\delta H_T  (\vec{p}_0) = \langle \psi_{\vec{p}_0, U}^{+}|H_T |\psi_{\vec{p}_0, L }^{+} \rangle + \textrm{H.c.}\,.
\end{align}
Using the effective Hamiltonian ($\delta H^{K}_{U/L} (\vec{p}_0)  + \delta H_T  (\vec{p}_0) $), we compute the localization length $\xi_{\parallel}$ along the straight line, where $\xi_{\parallel}$ is defined as the inverse of the projected inverse participation ratio on the line.
We show $\xi_{\parallel}$ as a function of the angle $\phi$ in Fig.~\ref{Fig:Incommensuratelocalization}~b).

Interestingly, for commensurate angles $\phi$ -- as, e.g., found at multiples of $\pi / 6$ -- the localization length exhibits a sharp peak, clearly showing that localization occurs due to incommensuration effects of the underlying \moire lattice. 
Whenever a wave function crosses the region of an incommensurate angle it gets tied up there, leading to localization. Fig.~\ref{Fig:Incommensuratelocalization}~c) shows that the localization length tentatively increases linearly with increasing $t_{\perp}$, however, with an additional oscillation. The oscillation effect effect occurs as a function of angle and we did not resolve its origin, here.\medskip

\begin{figure}[t!]
	\centering
	\includegraphics[width=\columnwidth]{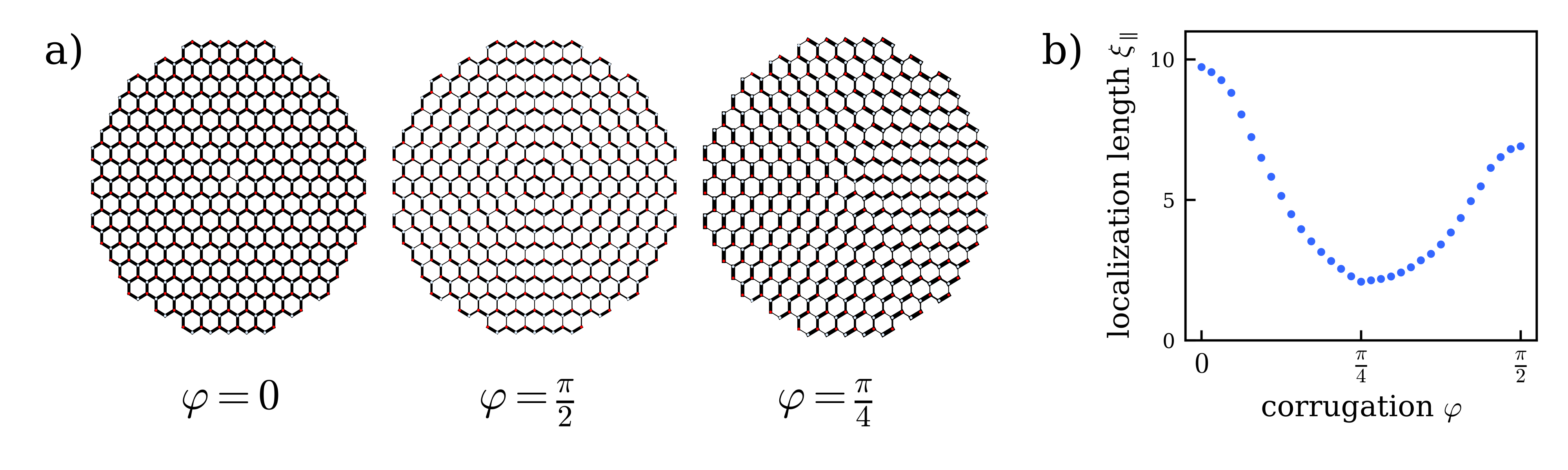}
	\caption{{\bf Corrugation effect.} (a) Diagrams to show how lattice sites in the momentum space are connected with the nearest neighbor sites by the interlayer hopping with a different level of the corrugation. The corrugation is effectively parameterized with $\varphi$ as $t^{\rm AA} = t_{\perp} \cos \varphi$ and $t^{\rm AB} = t_{\perp} \sin \varphi$. With the eigenstates $|\vec{k} + \vec{Q}_n, \alpha \rangle$ of the Hamiltonian in the absence of the interlayer coupling, the strength of the matrix element $ |\langle \vec{k}+ \vec{Q}_n^{'}, U |H_T|\vec{k} + \vec{Q}_n, L \rangle|$ is represented as the thickness of the connection between sites.  (b) localization length  $\xi_{\parallel}$ as a function of $\varphi$ with fixed $t_{\perp} = 0.5 v_F q_m$.}
	\label{Fig:Corrugation}
\end{figure}

\section{Corrugation effect}
\label{app:corrugation}

Finally, we show that in the weak corrugation case, an additional localization mechanism occurs along the energy circle. Here we address the effect of the corrugation within the momentum space continuum model.
This effect is due to interference between the interlayer couplings in the regions of AA stacking and AB stacking.
We note that in the main text, we use a strong corrugation case, i.e. $t^{\rm AA} = 0$ in Eq.~\eqref{hoppingFinal}, to avoid this effect as it is not of primary interest for the present work.

The corrugation can effectively be taken into account in the interlayer coupling terms of the continuum model, cf.~Eq.~\eqref{hoppingFinal}.
Using an angular parameter $\varphi$, it can be continuously tuned from weak to strong as $t^{\rm AA} = t_{\perp} \cos \varphi$ and $t^{\rm AB} = t_{\perp} \sin \varphi$.

To see how neighboring lattice points are connected by the interlayer coupling, we employ a perturbative approach.  First, we calculate the eigenvectors of the Hamiltonian in the absence of the interlayer coupling $|\vec{k} + \vec{Q}_n, \alpha = U, L \rangle$. 
We then compute matrix element
$ \langle \vec{k}+ \vec{Q}_n^{'}, U |H_T|\vec{k} + \vec{Q}_n, L \rangle$.
The absolute value of the matrix element is shown in Fig.~\ref{Fig:Corrugation}~a) and is represented by the thickness of the connections between lattice points.

Without corrugation ($\varphi = \pi /4$), the connections in the angular direction are suppressed due to the formation of dimer states in comparison with the connections in the radial direction. 
On the other hand, with strong corrugation, i.e. $\varphi = 0$ and $\varphi = \pi /2$, the connection along a contour of equal energy is relatively strong. 
While three nearest neighbor connections are equivalent in the $\varphi = 0$ ($t^{\rm AB} =0$) case, the connection in  the $\varphi = \pi/2$ ($t^{\rm AA}=0$) dominantly occurs along the angular direction.
Strong dependence on the corrugation suggests an interference effect between the AA and AB interlayer coupling.

The above localization mechanism is further supported by a calculation of the localization length along a ring, Fig.~\ref{Fig:Corrugation}~b). For decreasing corrugation, i.e. closer to $\varphi = \pi/4$, the localization length decreases.
This clearly shows that there is an additional localization mechanism in the weak corrugation case.


\end{document}